\documentclass{article}

 \usepackage[preprint]{neurips_2026}

\usepackage[utf8]{inputenc} 
\usepackage[T1]{fontenc}    
\usepackage{hyperref}       
\usepackage{url}            
\usepackage{booktabs}       
\usepackage{amsfonts}       
\usepackage{nicefrac}       
\usepackage{microtype}      
\usepackage{xcolor}         
\usepackage{graphicx} 
\usepackage{amsmath}
\usepackage{booktabs}
\usepackage{array}
\usepackage[most]{tcolorbox}
\usepackage{tabularx}
\usepackage{longtable}
\usepackage{caption}
\usepackage{wrapfig}
\usepackage{bbm}
\usepackage{multirow}
\usepackage{multicol}
\usepackage{enumitem}
\usepackage{hyperref}

\title{PPI2Text: Captioning Protein-Protein Interactions with Coordinate-Aligned Pair-Map Decoding}

\author{
\textbf{Xiao Fei\textsuperscript{1},
Sarah Almeida Carneiro\textsuperscript{1}, Yang Zhang\textsuperscript{1}, Lawrence P. Petalidis\textsuperscript{3},} \\
\textbf{Achilleas Tsortos\textsuperscript{4}, 
Costas Bouyioukos\textsuperscript{5},
Michalis Vazirgiannis\textsuperscript{1,2}} \\
\textsuperscript{1}École Polytechnique, Institut Polytechnique de Paris, France \\
\textsuperscript{2}Mohamed bin Zayed University of Artificial Intelligence, United Arab Emirates \\
\textsuperscript{3}M42 Health, United Arab Emirates\\
\textsuperscript{4}Foundation for Research and Technology - Hellas, Greece\\
\textsuperscript{5}Université Paris Cité, France \\
\texttt{xiao.fei@polytechnique.edu} \\
\texttt{\{almeidacarneiro, mvazirg\}@lix.polytechnique.fr}
}

\begin{document}

\maketitle

\begin{abstract}

Protein-protein interaction (PPI) modeling has been widely studied as a binary or multi-label classification task. While emerging multimodal large language models (LLMs) can now describe single proteins, they remain unable to generate free-form descriptions of interactions between protein pairs. 
Moving beyond controlled vocabulary annotations, we propose to model PPI using free-text description, enabling richer expressiveness, improved interpretability, and better integration with literature knowledge base. 
We present PPI2Text, a multimodal LLM for free-form PPI captioning from amino acid sequences, that encodes each protein using ESM3 encoder, constructs a pair map from the two representations to capture interactions across all residue pairs, and autoregressively generates descriptions using a Qwen3 language decoder. 
We further introduce PaCo-RoPE, a coordinate-aligned positional encoding that aligns each axis of the pair grid with the residue positions of the corresponding protein. 
In addition, we release PPI2Text-Dataset, a 351k-pair corpus of free-form PPI descriptions aggregated from ten curated biological databases and further synthesized with Gemini under evidence-tiered prompting. 
PPI2Text consistently outperforms strong baselines across multiple ablation settings and evaluation protocols. It not only achieves higher scores on linguistic metrics against synthesized references, but also excels on factuality metrics, where an LLM-based judge evaluates outputs against raw biological evidence. 
\footnote{The source code of PPI2Text model is available at: \href{https://github.com/ColinFX/PPI2Text}{https://github.com/ColinFX/PPI2Text}.}
\footnote{The 351k corpus data of PPI2Text-Dataset is available at: \href{https://huggingface.co/datasets/xiao-fei/PPI2Text-Dataset}{https://huggingface.co/datasets/xiao-fei/PPI2Text-Dataset}.}

\end{abstract}

\section{Introduction}
\label{sec:intro}

Protein–protein interactions (PPIs) drive a wide range of cellular processes by allowing proteins to form complexes involved in metabolism, signaling, DNA transcription and replication, and immune responses (\cite{stites1997protein}). While interaction characterization can be obtained through experimental validation, such approaches are time- and resource-intensive and do not always yield details (\cite{rao2014protein}). To reduce these practical constraints, existing computational PPI approaches mostly relied on classification and regression modeling to extract evolutionary signals and encode interactions through conservation patterns. However, these representations struggle to capture higher-order and non-linear dependencies and are further limited by heterogeneous data sources. Meanwhile, despite the biological importance of PPIs, the field remains comparatively understudied relative to single-protein function prediction and annotation. Furthermore, a substantial fraction of biologically relevant PPIs remain undiscovered or uncharacterized. Consequently, continued research into robust and interpretable computational approaches for PPI characterization remains essential for advancing our understanding of complex cellular systems and improving biological discovery at scale.

Despite the availability of large-scale biological repositories such as IntAct and STRING, PPI-related information remains fragmented and inconsistently structured. As a result, current approaches often reduce interaction prediction to discrete labels or fixed relational structures, which limits their ability to capture how or why a biological feature interact within context. Protein–protein interaction captioning offers a complementary paradigm in which interactions are represented as coherent free-text descriptions, rather than being reduced to fixed categorical labels or numerical scores. By expressing interactions through flexible natural-language narratives, this approach enables the integration and interpretation of individual protein-specific features and emergent properties arising from their interaction.
 Although not all observed protein features imply causal or correlation relationships, some interactions are strongly condition-dependent and emerge only through complex chains of biological events (\cite{chindelevitch2012causal}). Nevertheless, PPI free-text captioning remains an emerging research area, and relatively few studies have explored the logical integration of rigid interaction classifications into interpretable natural-language descriptions, despite its potential to improve biological interpretability, hypothesis generation, and contextual understanding of molecular interactions. Further details are elaborated in Appendix~\ref{ap:freeT}.

To address these limitations, we propose a unified framework that enhances both representation and data quality by shifting from structured prediction paradigms toward a more expressive formulation of PPI understanding. Our approach is designed to better reflect the contextual and relational nature of protein interactions, enabling richer modeling of dependencies across heterogeneous biological sources. Our contributions are summarized as follows:

\begin{itemize}[noitemsep, topsep=0pt]
    \item \textbf{Free-text PPI modeling framework:} We introduce a novel task formulation to cast PPI modeling as human-readable description generation, enabling direct interpretation of interaction semantics beyond fixed annotation schemas. 
    \item \textbf{Large-scale PPI description dataset:} We develop a data synthesis pipeline to augment natural language descriptions for PPI to improve expressiveness and cross-dataset generalization. A 351k-pair high-quality description corpus is released.
    \item \textbf{Interaction-aware protein encoding:} We propose \textit{PPI2Text} as a unified architecture to capture residue-level relationships between proteins through pairwise interaction map modeling, with coordinate-aligned rotary positional encoding (\textit{PaCo-RoPE}) to align multiple multimodal components and to boost understanding. 
    \item \textbf{Strong performance on empirical evidence:} We demonstrate that the model outperforms various baselines on both linguistic metrics against synthesized text as well as factuality metrics against raw evidence. 
\end{itemize}

The paper is structured as follows. Section~\ref{sec:rw} reviews related work on protein-protein interaction methods. Section~\ref{sec:dataset} introduces our description synthesis pipeline and the dataset construction. Section~\ref{sec:model} covers PPI2Text model architecture design. Experimental setup and results are shown in Section~\ref{sec:experiments}. Section \ref{sec:limit} discusses limitations and Section \ref{sec:conclude} concludes. Additional materials are provided in the Appendix.

\section{Related Work}
\label{sec:rw}

\paragraph{Sequence-based single-protein description:}
Early protein modeling relied on task-specific sequence-based methods with fixed-label supervision for function and interaction prediction. More recent large language models generate free-text descriptions from sequences (\cite{taylor2022galactica,liu2024prott3, zhou2025decoding, fei2025prot2text}), translating embeddings into natural language. However, they are designed for single protein input, and are trained on single-protein data and general biochemical text without detailed grounding in interaction data, limiting their ability to capture interaction-specific semantics and relational context.

\paragraph{Deep learning models for PPI prediction:}
RNN-based models have been widely used for sequence representation learning, often within hybrid frameworks that integrate structural or biophysical features (\cite{alakus2021novel, zhou2022residue}). LSTM-based approaches further improve performance through regularization techniques (\cite{tsukiyama2021lstm, szymborski2022rapppid, deng2021hybrid}). More recently, attention-based methods and transformer architectures have advanced PPI prediction by capturing long-range dependencies and complex protein relationships (\cite{hu2024protein, hu2024improving, li2022sdnn, wu2024attentionep, mou2023transformer, kang2023hn}). However, these approaches often rely on handcrafted features or hybrid pipelines, which can introduce bias and limit scalability. Emerging tools for PPI 3D structure prediction, like the AlphaFold family \cite{evans2021protein, abramson2024accurate}), have become widely used resources. However, these methods currently serve as orthogonal tools to PPI captioning because they produce structural representations of interactions rather than natural-language descriptions.

\paragraph{Sequence-based LLMs for PPI prediction: }Recent LLM-based approaches model protein–protein interactions as sequence-based prediction tasks using pretrained embeddings and prompting strategies (\cite{jin2024prollm, hallee2023protein}). Multimodal and retrieval-augmented methods incorporate sequence, text, and network data or biomedical evidence for improved prediction and reasoning (\cite{zhuo2024protllm, zhou2025large, ullanat2026learning, jeon2026ragppi}). However, these methods remain schema-bounded, producing labels or scores that compress biological context and limit the representation of mechanisms and evidence.

\section{Evidence-Tiered PPI Description Dataset Generation}
\label{sec:dataset}

\begin{table*}[t!] 
\centering
\begin{tcolorbox}[
  colback=gray!5,
  colframe=black,
  boxrule=0.8pt,
  arc=2mm,
  width=\linewidth
]
\small
\textbf{Raw Evidence:} \\[2pt]
[IntAct] \textcolor{orange}{\texttt{18 evidence lines, MIscore 0.94}}; physical association... \\[2pt]
[PubMed] \textcolor{orange}{\texttt{11 publications}} \\[2pt]
[UniProt] \textcolor{orange}{\texttt{ProtA:}} transcriptional regulator and general inhibitor of RNA polymerase II... \\[2pt]
[UniProt] \textcolor{orange}{\texttt{ProtB:}} catalytic subunit of P-TEFb; phosphorylates C-terminal domain of RNA Pol II... \\[2pt]
[CORUM] \textcolor{orange}{\texttt{2 complexes}}; 7SK RNP, P-TEFb-7SKRNA-HEXIM1 \\[2pt]
[STRING] \textcolor{orange}{\texttt{0.99 hybrid confidence}} \quad 
[3did] \textcolor{orange}{\texttt{null}} \quad [SIGNOR] \textcolor{orange}{\texttt{null}} \quad [ComplexPortal] \textcolor{orange}{\texttt{null}}

\tcbline

\textbf{Tags:} [Length] \textcolor{orange}{\texttt{long}} \quad [Verb] \textcolor{orange}{\texttt{firm}} \quad [Mechanism] \textcolor{orange}{\texttt{absent}} \quad [Silence] \textcolor{orange}{\texttt{none}}

\tcbline

\textbf{Synthesis: }Protein A and Protein B are core components of the 7SK RNP complex and the P-TEFb-7SKRNA-HEXIM1 complex, where they physically associate to regulate RNA polymerase II transcription. Protein A is a transcriptional regulator belonging to the HEXIM family... \\[2pt]
The catalytic activity of Protein B is distributed throughout its protein kinase domain and is regulated by a complex pattern of post-translational modifications, including autophosphorylation at the T-loop, N-terminal acetylation, and polyubiquitination by UBR5...
\end{tcolorbox}

\noindent\begin{minipage}{\textwidth}
\captionof{figure}{An example of the dataset generation (interaction between HEXIM1 and CDK9). The top section shows the raw data aggregated from multiple heterogeneous sources, while the middle section lists the control tags computed from the aggregated evidence profile, and the bottom section is the labeler-LLM output, synthesized from the raw evidence under the control-tag constraints. }\label{fig:augmentation_example}
\end{minipage}

\label{tab:aug_ex}
\end{table*}

PPI annotations capture rich biological knowledge but remain sparse, heterogeneous, and fragmented. Integration efforts such as OmniPath (\cite{turei2016omnipath}) and ConsensusPathDB (\cite{kamburov2009consensuspathdb}) unify these data, but rely on rigid schemas that limit contextual expressivity and evidence integration. In this paper, we propose a dataset synthesis labeling pipeline that projects PPI data into a natural language space via free-form text augmentation, enabling more expressive interaction modeling and better cross-dataset generalization.

\paragraph{Source aggregation: }
We first construct a comprehensive and reliable factual base by integrating ten complementary sources capturing interaction-level evidence, molecular context, structural constraints, and higher-order functional organization: IntAct (\cite{kerrien2012intact}), PubMed (\cite{canese2013pubmed}), UniProt (\cite{uniprot2019uniprot}), 3did (\cite{mosca20143did}), Pfam (\cite{mistry2021pfam}), STRING (\cite{mering2003string}), SIGNOR (\cite{perfetto2016signor}), Reactome (\cite{croft2010reactome}), CORUM (\cite{giurgiu2019corum}), and ComplexPortal (\cite{meldal2015complex}). These sources are selected to ensure coverage across experimental, structural, and curated biological evidence, with further details provided in Appendix~\ref{apd:10datasets}.

After collection, integration and de-duplication, we obtain approximately 1.08M protein pairs. However, annotation quality is highly imbalanced. Many PPIs are supported only by single high-throughput assays (e.g., yeast two-hybrid), which provide limited mechanistic resolution. This heterogeneity introduces a fundamental trade-off between descriptive expressivity and factual fidelity in downstream synthesis: while strongly supported PPIs allow reliable detailed descriptions, weakly supported interactions risk hallucinated or over-specified narratives.

\paragraph{Evidence-tiered quality filtering:} To address this, we introduce an evidence-tiered filtering and control strategy. We define a unified \textit{evidence score} $E(r)$ for each pair $r$ that integrates complementary interaction-level and contextual biological signals, capturing both direct experimental support and broader biological coherence between interacting partners. We aggregate these signals while respecting their structured dependencies and heterogeneous reliability, yielding a score that reflects consistency, redundancy, and overall evidential strength. Further details can be found in Appendix~\ref{apd:evidence_score}.

We then apply K-means clustering over the evidence scores to partition PPIs into three tiers (T1–T3), ranging from weakest to strongest support. We discard T1 due to insufficient evidence beyond single high-throughput assays, retain all T3 interactions as strongly curated, and apply a homology-aware subsampling strategy to intermediate T2 tier using MMSeq2 (\cite{steinegger2017mmseqs2}), selecting representative high-evidence pairs within homologous clusters. This process yields approximately 351k high-confidence PPIs, balancing diversity with evidential reliability and forming a robust foundation for LLM-based synthesis. Further details are provided in Appendix~\ref{apd:kmeans} and \ref{apd:bias}.

\paragraph{Description synthesis pipeline:}
Finally, we augment the retained PPIs using a labeler LLM. The full system prompt is detailed in Appendix~\ref{ap:synthesisPrompt}. To ensure faithfulness to the underlying evidence, we introduce a tier-controlled generation framework that explicitly constrains the output space according to the strength of biological support. This reduces the effective hypothesis space of the model, preventing unsupported extrapolation or over-detailed descriptions for weakly supported interactions. Four orthogonal constraints are explicitly enforced in each prompt: \textit{descriptive granularity}, \textit{epistemic strength}, \textit{mechanistic attribution}, and \textit{silence policy} (see Appendix \ref{apd:tags}).

We release the resulting dataset of 351k protein interactions, each paired with a free-form textual description of its interaction context and mechanism. We evaluate generation quality across Gemini-3-Pro-Preview (\cite{team2023gemini}), Claude-Opus-4.6 (\cite{bai2022constitutional}), and GPT-5.4 (\cite{radford2018improving}), focusing on factual correctness, lexical quality, and linguistic fluency. Among these, Gemini consistently achieves better expert-rated biological fidelity and overall output quality.

\section{Coordinate-Aligned Pair-Map Decoding}
\label{sec:model}

\begin{figure} [t!]
    \centering
    \includegraphics[width=\linewidth]{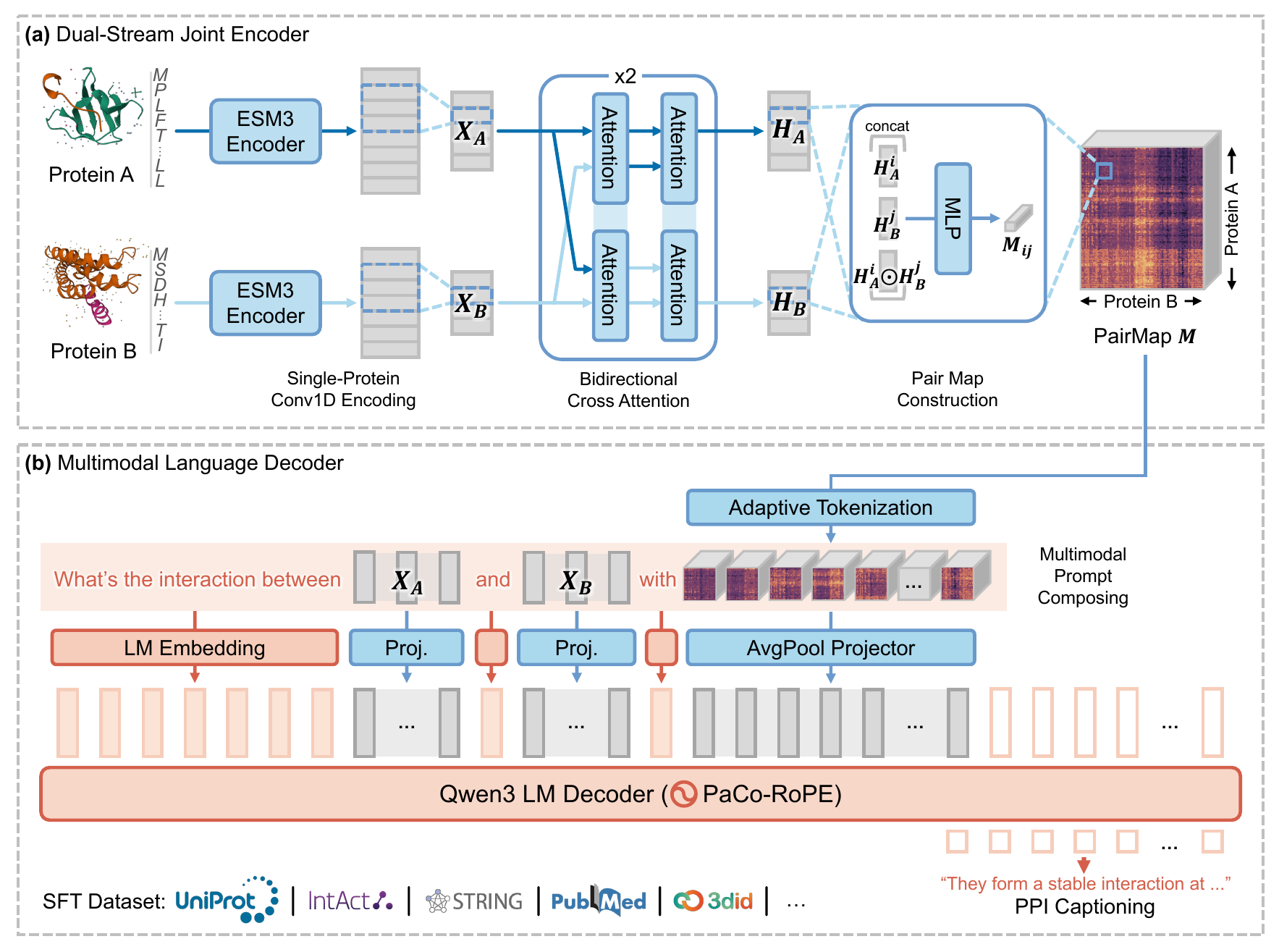}
    \caption{Model architecture of PPI2Text. \textbf{(a)} In dual-stream joint encoder, both proteins are first separately encoded and compressed in parallel then jointly encoded to construct a PairMap, modeling interactions between pairs of residues. \textbf{(b)} For multimodal language decoding, both single-protein representations and pair map tokens from the encoder are projected then concatenated with text token embeddings to compose the interleaved multimodal prompt, and finally a Qwen3 language model decoder with Pair-Coordinated RoPE (PaCo-RoPE) generates the free-text description of the interaction.}
    \label{fig:architecture}
\end{figure}

PPI takes place at the interface between the two proteins and is determined by the set of contacts between their residues. A natural representation of this information is a residue-level coupling matrix, with positions on protein A along one axis and positions on protein B along the other (\cite{evans2021protein}). In this paper, we propose \textit{PPI2Text}, as shown in Figure~\ref{fig:architecture}, that models the interaction by this 2D representation, then bridges it to the 1D token sequence of a language decoder, enabling direct attention over pairwise residue interactions. The design avoids compressing the 2D interaction map into a global vector or relying solely on single-protein embeddings.

\subsection{Single-Protein Parallel Encoding}
\label{sec:method:encoder}

We use pretrained ESM3 (\cite{doi:10.1126/science.ads0018}) as the frozen single-protein encoding backbone. ESM3 is a multimodal foundation model jointly pretrained over multiple per-residue tracks including amino-acid sequence, AlphaFold2-predicted (\cite{jumper2021highly}) 3D structure coordinates, plus the derived discretized structure tokens, secondary structure (SS8), solubility factors (SASA), and structure confidence scores (pLDDT). For each protein $P \in \{A, B\}$, it returns per-residue embeddings $E_P \in \mathbb{R}^{L_P \times d_{ESM}}$, where $L_P$ is the length of the protein (see Appendix \ref{ap:ESM3}). We use ESM3 specifically because its multitrack pretraining already entangles evolutionary and structural signals in a single tensor.

Then, a two-layer strided 1D convolutional compressor reduces the sequence-long dimension of the embedding $X_P$ by a factor of four, with parameters shared across the two proteins, so that $X_P\in\mathbb{R}^{L_P'\times d_{hidden}}$, where the compressed length $L'_P = \lfloor L_P / 4 \rfloor$ depends only on $L_P$. Since per-residue ESM3 embeddings are already context-rich through the encoder's full self-attention stack, a moderate pooling preserves most of the local information. Moreover, compression is necessary because both protein sequences can be up to thousands of residues long, and uncompressed representations can lead to redundant and noisy inputs to the decoder, affecting both training and inference efficiency. The two single-protein representations $X_P$ are projected then passed as part of the input embeddings to the decoder to provide partner context information on the interaction. However, we argue that the interaction information cannot be fully captured by single-protein information, and a 2D pair map is necessary. 

\subsection{Pair-Map Encoding}
\label{sec:method:pairmap}

We use two bidirectional cross-attention blocks to exchange information between the two protein representations before any pair representation is formed, with weights also shared between the directions $A \to B$ and $B \to A$. Each block combines a cross-attention to the partner and a self-attention to each protein itself, and the output of the last block is marked as $H_A \in \mathbb{R}^{L'_A \times d_c}$ and $H_B \in \mathbb{R}^{L'_B \times d_c}$. As inspired by MINT (\cite{ullanat2026learning}), the cross attention mechanism naturally guide corresponding interactive segments to enrich then highlight their role and context in the interaction. Furthermore, we interleave self-attention layers to stabilize the interaction embedding and reinforce the integration of global contextual information for each protein. 

Then, for each $(i, j) \in [L'_A] \times [L'_B]$, we form a per-position pair feature
\begin{equation}
M_{ij} = \phi_\mathrm{pair}\!\left([H_A^i \,\Vert\, H_B^j \,\Vert\, H_A^i \odot H_B^j]\right) \in \mathbb{R}^{d_{hidden}},
\label{eq:pairmap}
\end{equation}
where $\Vert$ is concatenation, $\odot$ is the Hadamard product, and $\phi_\mathrm{pair}: \mathbb{R}^{3 d_{hidden}} \to \mathbb{R}^{d_{hidden}}$ is a 2-layer MLP with GeLU as the activation function of the first layer to introduce non-linearity. The Hadamard term captures coordinate-wise feature co-activation between $H_A^i$ and $H_B^j$. Remarkably, we choose to project every pair of $H_A^i$ and $H_B^j$ into an embedding vector instead of a scalar, aiming to maximize the semantic understanding of the mechanistic details and biological context. The resulting 3D tensor $M \in \mathbb{R}^{L'_A \times L'_B \times d_{hidden}}$ is thus length-dependent in this case.

To inject the pair-map into the decoder as pseudo-tokens, we apply adaptive average pooling to a target $H_t \times W_t$ grid, flatten the grid into sequential patches, and project the result to the decoder's hidden space:
\begin{align}
\bar{M} &= \mathrm{AdaptAvgPool}_{H_t \times W_t}(M), \\
U &= \mathrm{RMSNorm}(\phi_\mathrm{tok}(\bar{M})) \in \mathbb{R}^{H_t W_t \times d_g},
\label{eq:pairtokens}
\end{align}
where $\phi_\mathrm{tok}$ is a 2-layer MLP and $d_{qwen}$ is the Qwen3 hidden size. We set $H_t = W_t = 32$, which adds 1024 pair-map tokens regardless of protein length. The fixed token budget in the decoder aims to maintain a rational information density and local granularity so that the decoder can focus on most significant aspects of the interaction without being distracted to noisy side-elements. The adaptive pooling also preserves the variety of information in each patch by orthogonal elements in the higher dimension representations. A Root-Mean-Square (RMS) normalization is applied in the end to align the scale of $U$ with the text token embeddings.

\subsection{Coordinate-Aligned Decoding}

In the end, both single-protein representations $X_P$ and the tokenized pair map representations $U$ are inserted into the natural language prompt sentence with their corresponding interleaved positions. The composed multimodal prompt is passed to Qwen3 language model decoder which auto-regressively generates the description of the interaction. 

\begin{figure} [t!]
    \centering
    \includegraphics[width=0.85\linewidth]{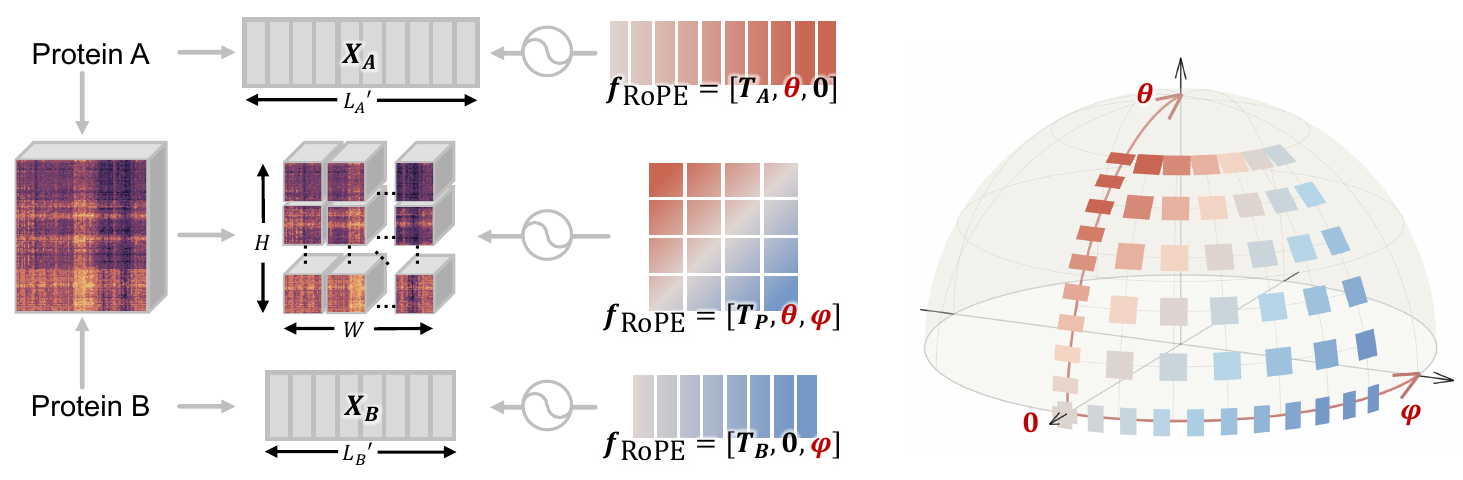}
    \caption{Illustration of the PaCo-RoPE as an extension to standard rotary positional embedding to paired-protein inputs. Each single protein representations $X_A$ , $X_B$ carries its own positional axis, and the two axes of the pair map are inherited from the single-protein axes correspondingly with residue-level alignment.}
    \label{fig:pacorope}
\end{figure}

However in the pretrained Qwen3 language model (\cite{qwen3}), the native 1D positional encoding over the flattened prompt would give the 1024 pair-map tokens consecutive indices, discarding the 2D grid structure of the pair-map and decoupling the pair-map tokens from the residue indices of the two proteins. We propose \textit{PaCo-RoPE} to address both issues. It extends the multimodal interleaved RoPE of Qwen3-VL (\cite{bai2025qwen3}) to a pair-coordinate grid, where each protein $P\in\{A,B\}$ is tied to a dedicated spatial channel and the pair-map sits at the intersection of the two protein channels. 

As illustrated in the Figure \ref{fig:pacorope}, every text or multimodal token is mapped to a unique 3D position IDs $(p^T, p^\theta, p^\varphi)$ as follows: 

\textbf{Text tokens}: For text token at position $n$ in the full sequence of prompt plus the generated answer, we apply:
    \begin{equation}
    (p^T, p^\theta, p^\varphi) = (n,n,n)
    \end{equation}
We collapse the 3D-RoPE to its canonical 1D form where all three channels are filled with their actual positions in the sequence. The behavior falls back to the native settings of Qwen3 model, inheriting the pretrained natural language ability.

\textbf{Single-protein tokens (protein A)}: For single-protein representation tokens of protein A ($X_A$) that is at position $n$ in the full sequence but represents continual residues centered at position $i$ in the protein A, we apply:
    \begin{equation}
        (p^T, p^\theta, p^\varphi)=(n, \lfloor i / 4 \rfloor, 0)
    \end{equation}
We specially use the second channel to indicate the location of the residue in the amino-acid sequence of protein A, while the third channel is muted. The first channel is maintained to represent the actual position in the full sequence. Such orthogonal isolation treats the protein sequence A as an independent modality to the text, enabling better understanding by the model.

\textbf{Single-protein tokens (protein B)}: Similarly, for single-protein representation tokens of protein B ($X_B$), we use the third channel for $j$ the centered location of the residue in the protein B while the second channel is muted instead: 
    \begin{equation}
        (p^T, p^\theta, p^\varphi)=(n,0,\lfloor j / 4 \rfloor)
    \end{equation}
As we isolate the position dimension of these two partners for the interaction, the decoder treats them independently without confusion on the dual multimodal inputs. 

\textbf{Pair-map tokens}: For pair map patch of grid $(m,n)$ and serialized into token at position $n$ in the full sequence, we apply: 
    \begin{equation}
        (p^T, p^\theta, p^\varphi)=\left(n,(m+\frac{1}{2})\frac{L_A'}{H}, (n+\frac{1}{2})\frac{L_B'}{W}\right)
    \end{equation}
so that their grid positions are tied to the position of residues on both protein A and B. Due to the adaptive average pooling that segment the entire pair map into small patches, the hop in distance on neighboring single protein embeddings is different to that on neighboring patches. With both height and width coordinates aligned between the pair map and both single proteins, the decoder can not only better understands the 2D nature of the pair map despite the serialized tokenization, but also better process these two different forms of multimodal inputs comprehensively. 


After all, the encoding supplies a coordinate-aligned shortcut between the pair map and each per-protein stream at zero parameter cost, and preserves the maximal pretrained behavior of the Qwen3 decoder (see Appendix \ref{ap:PaCo-R} for further details).

\subsection{Training Setup}
\label{sec:method:training}

We fine-tune the PPI2Text model in an end-to-end way with cross-entropy loss computed only on response tokens under a teacher-forcing protocol. Both pretrained ESM3 and pretrained Qwen3 weights stay frozen, and we apply LoRA (\cite{hu2022lora}) to Qwen3, which closes the residual gap between the pretrained text distribution and the PPI description manifold. The remaining new modules are trained from scratch with a separate optimizer group, since pretrained adapters and from-scratch modules require different update scales.

\section{Experiments and Results}
\label{sec:experiments}

\paragraph{Dataset and Test Settings:} 
We evaluate our model on the synthesized PPI description dataset introduced in Section~\ref{sec:dataset}, focusing on generalization under two complementary test settings that capture different notions of unseen proteins, ranging from realistic deployment conditions to strict out-of-distribution (OOD) generalization.
First, we propose a \textit{temporal holdout} split based on chronological partitioning where all PPIs annotated after May 1, 2025 are reserved for evaluation. This setting approximates a realistic deployment scenario in which the model predicts newly discovered interactions using previously available biological knowledge. To further prevent leakage, pair-level similarity-based decontamination is applied to the remaining training samples.
Second, we increase difficulty by using a \textit{C3-hard} split that enforces strict protein-level decontamination, ensuring that no homologous proteins are shared between training and test sets. As a result, all proteins similar to those seen during training are removed from the test set, simulating the discovery of entirely novel protein interactions. This setting is substantially more challenging than a standard holdout split because it requires the model to generalize under completely unseen biological contexts. This stress-test setting follows the increasingly adopted C3 protocol and the evaluation framework of \cite{10.1093/bib/bbae076}, assessing the model’s OOD generalization to entirely unseen proteins.
The model is trained on 280k PPIs, with a validation set of 2,500 samples. The temporal holdout and C3-hard test sets contain 1,690 and 2,730 interactions, respectively. Further details about the decontamination are provided in Appendix~\ref{ap:SplittingProtocol}.

  \begin{table}[!t]
  \small
  \centering
  \caption{Baseline and ablation results on temporal holdout set. Lexical metrics includes BLEU-2/4 F1 scores (B-2/4), ROUGE-1/2/L F1 scores (R-1/2/L), with semantic BERTScore using RoBERTa (RBT) and BioBERT (BBT) embeddings. LLM-as-a-judge scores against the raw evidence cover Entities (Ent), Interaction (Int), Mechanism (Mec), and the Average (Avg).}
  \resizebox{\textwidth}{!}{%
  \begin{tabular}{l!{\vrule width 1pt}cc!{\vrule width 0.5pt}ccc!{\vrule width 0.5pt}cc!{\vrule width 1pt}ccc!{\vrule
   width 0.5pt}c}

  \toprule

  & \multicolumn{7}{c!{\vrule width 1pt}}{\textbf{(against Synthesized Text)}} &
  \multicolumn{4}{c}{\textbf{(against Raw Evidence)}} \\
  [4pt]

  \textbf{Model} & \textbf{B-2} & \textbf{B-4} & \textbf{R-1} & \textbf{R-2} & \textbf{R-L} & \textbf{RBT} & \textbf{BBT} &
  \textbf{Ent} & \textbf{Int} & \textbf{Mec} & \textbf{Avg} \\

  \midrule

  Seq+Qwen3            & 37.37          & 20.87          & 52.83          & 23.14          & 28.66
    & 88.02          & 80.74          & 1.94          & 7.50          & 1.85          & 3.76          \\
  MINT+Qwen3                    & 41.54          & 25.36          & 56.22          & 29.17          & 33.67
    & 89.03          & 82.48          & 2.68          & 7.80          & 1.63          & 4.04          \\
  \midrule
  SingProt-Only PPI2Text              & 43.35          & 27.61          & 58.44          & 31.88          & 36.00
    & 89.54          & 83.71          & 4.75          & 8.26          & 2.61          & 5.21          \\
  PairMap-Only PPI2Text           & 40.98          & 25.36          & 56.96          & 29.79          & 34.02
    & 89.23          & 83.09          & 3.22          & 7.95          & 1.03          & 4.07          \\
  1D-RoPE PPI2Text                & 39.17          & 23.61          & 54.63          & 27.45          & 32.50
    & 88.52          & 81.66          & 3.00          & 7.31          & 2.01          & 4.11          \\
  No-Cross PPI2Text               & 45.61          & 29.86          & 60.24          & 34.71          & 38.91
    & 89.94          & 84.02          & 4.94          & 7.98          & 4.67          & 5.86          \\
  \midrule
  \textbf{PPI2Text}               & \textbf{48.33} & \textbf{31.86} & \textbf{62.15} & \textbf{36.56} &
  \textbf{39.96} & \textbf{90.79} & \textbf{86.33} & \textbf{7.72} & \textbf{8.32} & \textbf{5.59} & \textbf{7.21} \\
  \bottomrule
  \end{tabular}
  }
  \label{tab:quant-ppi2text-hold}
  \end{table}

\paragraph{Baselines and Ablations:} As prior works focus on structured label prediction that are not directly comparable in output space, we evaluate PPI2Text through representative learning backbone baselines and ablations. \textit{Seq+Qwen3} uses no protein-specific encoder, but feeds raw amino-acid sequences directly to the decoder as trainable special tokens. \textit{MINT+Qwen3} employs the pretrained state-of-the-art PPI encoder MINT with Qwen3 as the decoder. We further design four ablation variants to analyze the contributions of individual components. \textit{SingProt-Only} passes only the single-protein representations $X_A$ and $X_B$ to the decoder, while \textit{PairMap-Only} uses only the tokenized pair map $U$. The role of the proposed PaCo-RoPE is examined in \textit{1D-RoPE}, where it is downgraded to the standard encoding in Qwen3. Finally, \textit{No-Cross} removes the bidirectional attention mask module to evaluate its impact relative to the full \textit{PPI2Text} model. For further reproducibility information on hyperparameters and computational resources, see Appendix \ref{apd:hparams}.

\paragraph{Evaluation Metrics:} We first evaluate the prediction from PPI2Text against the synthesized text of test samples. We use lexical metrics including BLEU scores and ROUGE scores, then semantic metrics like BERTScore with RoBERTa and BioBERT embeddings. Moreover, to decouple any potential information bottleneck or hallucination in the synthesized text, we evaluate the prediction against the raw evidence that are gathered from sources before the synthesis. To scale up the evaluation, we use Claude-Opus-4.7 as a judge to evaluate three orthogonal aspects of the generated test. Raw evidence evaluates directly extracted interaction components: Entities (Ent), Interaction (Int), Mechanism (Mec), and the metric's overall average (Avg). 

\begin{figure} [t!]
    \centering    
    \includegraphics[width=0.9\linewidth]{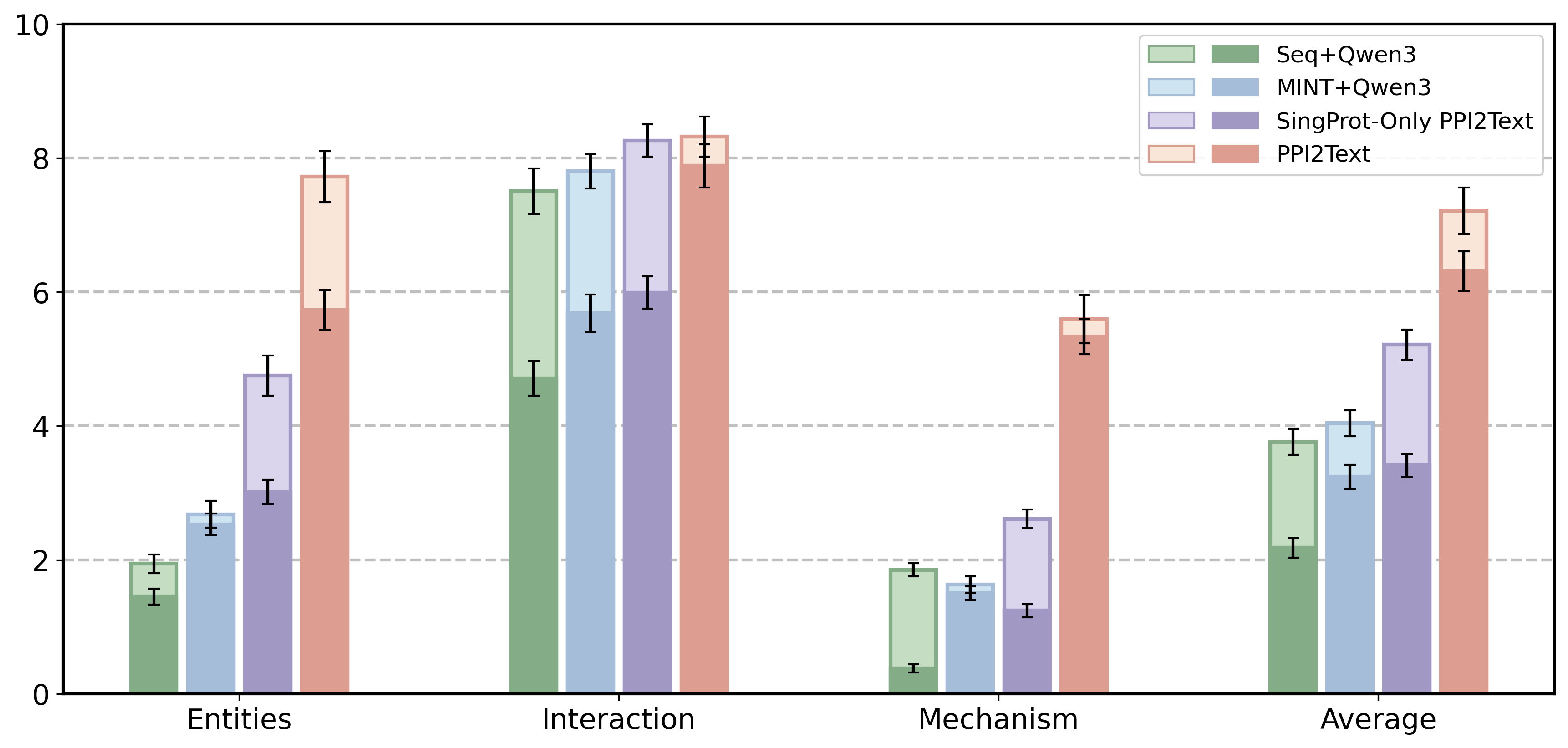}
    \caption{LLM-as-a-judge scores on baselines and ablations: evaluating the factual correctness of predicted text against raw evidence. Lighter bars denote performance on the temporal holdout split, while darker colors denote performance on the C3-hard split. PPI2Text consistently outperforms all other methods by a clear margin.}
    \label{fig:graphs}
\end{figure}

\paragraph{Results:}

Through our proposed evaluation setup, the performance of baseline models and PPI2Text is reported in Table \ref{tab:quant-ppi2text-hold}. As expected, the raw-sequence model without protein encoder performs badly, highlighting the limited biological signal available when relying solely on amino acid tokens. Meanwhile, models leveraging pretrained encoders such as MINT witnesses a slight improvement, likely because they capture richer evolutionary, sequential, and structural information learned from large-scale protein data. The ablation study further clarifies the role of each component in our framework. The SingProt-Only variant already provides noticeable improvements over encoder-free baselines, yet along with the PairMap-Only variant, we show that removing either single protein representations or pairmap tokens affects the performance. This is likely because these two components support each other and help the decoder to better understand the full picture of the interaction. The 1D-RoPE variant performs badly, suggesting that flattening the pair map and applying 1D positional encoding disrupts its inherent 2D structure, highlighting the necessity of our PaCo-RoPE design. Meanwhile, the No-Cross variant performs worse compared to the main proposal, indicating that the bidirectional cross attention blocks are important in the construction of the pairmap. 
These observations are further supported by the raw evidence-based evaluation. PPI2Text achieves the strongest overall performance, with particularly strong results in entity grounding and mechanism fidelity, indicating that the model not only produces fluent descriptions but also preserves key biological entities and interaction mechanisms, yielding more reliable and biologically meaningful predictions.

Figure \ref{fig:graphs} compares performance across the temporal holdout and C3-hard splits (see Appendix \ref{ap:raw_eval_prompt} and \ref{ap:SplittingProtocol}). As expected, the temporal holdout is more accessible, with all models showing higher and more similar performance due to overlap in training and test distributions. In contrast, the C3-hard split is substantially more challenging, as no homologous proteins are shared between splits. We can see a metric decrease in all of the compared propositions. Under this challenging setting, PPI2Text still consistently outperforms both baselines and ablations. Importantly, these gains are also reflected in raw evidence-based metrics, suggesting they stem from improved biological grounding rather than surface-level learning. Overall, PPI2Text demonstrates robust generalization in free-text PPI modeling, particularly under out-of-distribution conditions, with automated evaluation aligning with human expert assessments.

\section{Limitations and Future Work}
\label{sec:limit}

A limitation of the framework is its assumption that every input pair forms a protein–protein interaction, requiring a prior PPI screening step, although binary PPI screening is a well-studied task with many strong existing models. In addition, the lack of wet-lab validation limits empirical verification of de-novo predictions. Future work will incorporate interaction-level structural information, including predicted complex structures from models such as AlphaFold3, beyond the individual protein structures currently used. Additional safeguards are discussed in Appendix \ref{sec:safe}.

\section{Conclusion}
\label{sec:conclude}

In this work, we introduce, to our knowledge, the first framework for generating free-form, human-readable descriptions of protein–protein interactions. By moving beyond rigid annotation schemas and task-specific classifiers, our approach models interaction semantics in natural language, enabling richer and more interpretable representations. Experiments with strong baselines and targeted ablations show that each component contributes meaningfully, and that the proposed interaction-aware encoding captures fine-grained residue-level relationships. These improvements are reflected in both standard text generation metrics and evidence-grounded evaluations, indicating greater biological consistency and mechanistic fidelity. We believe this work also establishes a foundation for future research opening directions for downstream applications such as knowledge extraction and reasoning. To support further progress, we release a dataset of 315k interaction descriptions to encourage the development of more expressive and interpretable models.

\section{Acknowledgement}

This work was granted access to the HPC resources of IDRIS made by GENCI.

\bibliographystyle{plainnat}
\bibliography{refs}

\newpage


\appendix
\section*{Appendix}

\section{Why Free-Text Supervision Matters for PPI Modeling: Example Cases}
\label{ap:freeT}

Despite the extensive body of work proposing classification and regression-based approaches for protein–protein interaction (PPI) modeling, our free-text supervision framework offers a fundamentally different perspective. Traditional methods typically rely on structured outputs, such as binary interaction labels, categorical annotations, or predefined graph triples, which constrain the learning objective to a limited representation space. However, PPI modeling is inherently a generative downstream task. As a result, enforcing structured targets introduces a mismatch between the training objective and the actual output space of the model, potentially limiting expressiveness and fidelity.

\newcommand{\synth}[1]{\textcolor{orange}{#1}}

\tcbset{ppiexamplebox/.style={
  colback=gray!5, colframe=black, boxrule=0.8pt, arc=2mm, width=\linewidth,
  colbacktitle=gray!15, coltitle=black, fonttitle=\small
}}

\begin{figure}[htbp]
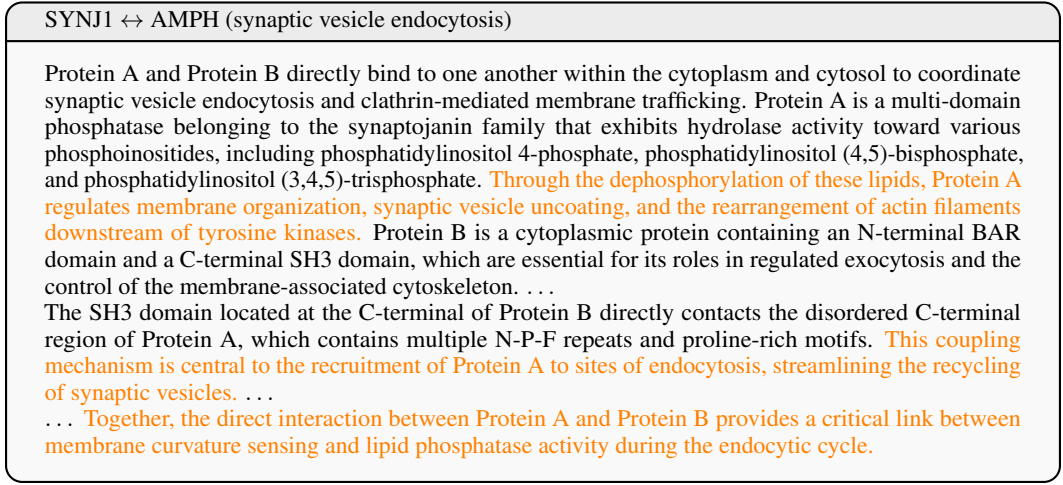

\centering
\begin{tcolorbox}[ppiexamplebox,
  title={SYNJ1 $\leftrightarrow$ AMPH (synaptic vesicle endocytosis)}]
\small
Protein A and Protein B directly bind to one another within the cytoplasm and cytosol to coordinate synaptic vesicle endocytosis and clathrin-mediated membrane trafficking. Protein A is a multi-domain phosphatase belonging to the synaptojanin family that exhibits hydrolase activity toward various phosphoinositides, including phosphatidylinositol 4-phosphate, phosphatidylinositol (4,5)-bisphosphate, and phosphatidylinositol (3,4,5)-trisphosphate. \synth{Through the dephosphorylation of these lipids, Protein A regulates membrane organization, synaptic vesicle uncoating, and the rearrangement of actin filaments downstream of tyrosine kinases.} Protein B is a cytoplasmic protein containing an N-terminal BAR domain and a C-terminal SH3 domain, which are essential for its roles in regulated exocytosis and the control of the membrane-associated cytoskeleton. $\dots$

The SH3 domain located at the C-terminal of Protein B directly contacts the disordered C-terminal region of Protein A, which contains multiple N-P-F repeats and proline-rich motifs. \synth{This coupling mechanism is central to the recruitment of Protein A to sites of endocytosis, streamlining the recycling of synaptic vesicles.} $\dots$

$\dots$ \synth{Together, the direct interaction between Protein A and Protein B provides a critical link between membrane curvature sensing and lipid phosphatase activity during the endocytic cycle.}
\end{tcolorbox}
\caption{Example of one-to-many causal effects plus cross-domain bridging. A single enzymatic activity (lipid dephosphorylation) is unfolded into three distinct downstream consequences and the closing sentence bridges two otherwise disjoint semantic domains, membrane geometry sensing and lipid phosphatase chemistry.}
\label{fig:synth_synj1_amph}
\end{figure}

\begin{figure}[htbp]
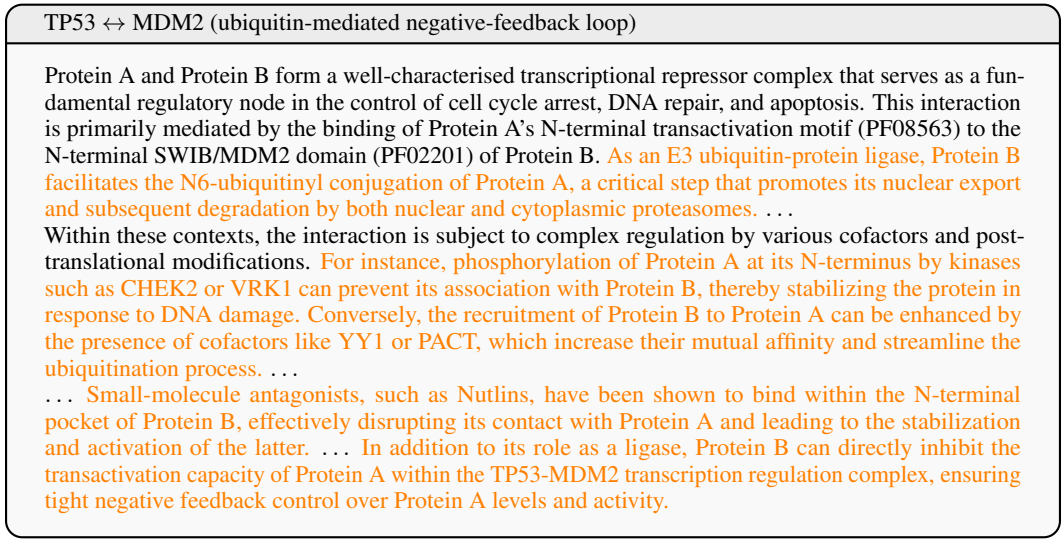

\centering
\begin{tcolorbox}[ppiexamplebox,
  title={TP53 $\leftrightarrow$ MDM2 (ubiquitin-mediated negative-feedback loop)}]
\small
Protein A and Protein B form a well-characterised transcriptional repressor complex that serves as a fundamental regulatory node in the control of cell cycle arrest, DNA repair, and apoptosis. This interaction is primarily mediated by the binding of Protein A's N-terminal transactivation motif (PF08563) to the N-terminal SWIB/MDM2 domain (PF02201) of Protein B. \synth{As an E3 ubiquitin-protein ligase, Protein B facilitates the N6-ubiquitinyl conjugation of Protein A, a critical step that promotes its nuclear export and subsequent degradation by both nuclear and cytoplasmic proteasomes.} $\dots$

Within these contexts, the interaction is subject to complex regulation by various cofactors and post-translational modifications. \synth{For instance, phosphorylation of Protein A at its N-terminus by kinases such as CHEK2 or VRK1 can prevent its association with Protein B, thereby stabilizing the protein in response to DNA damage.} \synth{Conversely, the recruitment of Protein B to Protein A can be enhanced by the presence of cofactors like YY1 or PACT, which increase their mutual affinity and streamline the ubiquitination process.} $\dots$

$\dots$ \synth{Small-molecule antagonists, such as Nutlins, have been shown to bind within the N-terminal pocket of Protein B, effectively disrupting its contact with Protein A and leading to the stabilization and activation of the latter.} $\dots$ \synth{In addition to its role as a ligase, Protein B can directly inhibit the transactivation capacity of Protein A within the TP53-MDM2 transcription regulation complex, ensuring tight negative feedback control over Protein A levels and activity.}
\end{tcolorbox}
\caption{Example of multi-regime contrastive reasoning. The same pair of proteins is described under three opposing regimes within one paragraph: CHK2/VRK1 phosphorylation prevents the binding, YY1/PACT cofactor recruitment enhances it, and Nutlin small molecules disrupt it.}
\label{fig:synth_tp53_mdm2}
\end{figure}

\begin{figure}[htbp]
\centering
\begin{tcolorbox}[ppiexamplebox,
  title={HRAS $\leftrightarrow$ RAF1 (nucleotide-state-dependent activation of the MAPK cascade)}]
\small
Protein A directly binds Protein B to form the HRAS-RAF1 complex, a foundational coupling in mitogenic signal transduction that up-regulates Protein B activity. \synth{The binding is strictly dependent on the nucleotide-bound state of Protein A, which functions as a small GTPase; it preferentially associates with Protein B when in its active GTP-bound conformation rather than its inactive GDP-bound state.} The N-terminal effector region of Protein A contacts the N-terminal Ras-binding domain (RBD) of Protein B. Additional structural evidence indicates that Protein A also contacts a phorbol-ester/diacylglycerol-binding C1 domain in the N-terminal region of Protein B. $\dots$

\synth{The assembly of the HRAS-RAF1 complex facilitates the translocation of Protein B to the plasma membrane, where it is activated through a complex process involving the release of autoinhibition, dimerization, and phosphorylation.} $\dots$ \synth{This pathway is subject to intricate feedback regulation; for instance, downstream kinases can phosphorylate Protein B at multiple sites in its N-terminal, central, and C-terminal regions to desensitize its activity.} $\dots$

$\dots$ Biochemical evidence further indicates that Protein A can be ubiquitinated by the BCR(LZTR1) E3 ubiquitin ligase complex at its C-terminal region, \synth{a modification that decreases its membrane association and inhibits downstream signaling.} $\dots$
\end{tcolorbox}
\caption{Example of state-gated activation cascade. The binding event triggers an ordered chain of physical state transitions: membrane translocation, autoinhibition release, dimerisation, and eventually phosphorylation. Each step is conditional on the previous, and the binding itself is gated by the GTP- vs.\ GDP-bound state of HRAS. }
\label{fig:synth_hras_raf1}
\end{figure}

\begin{figure}[htbp]
\centering
\begin{tcolorbox}[ppiexamplebox,
  title={HSPA8 $\leftrightarrow$ BAG1 (allosteric coupling in the Hsp70 nucleotide-exchange cycle)}]
\small
Protein A and Protein B directly bind to form a heteromeric complex that facilitates nucleotide exchange within the heat shock protein 70 chaperone cycle. This interaction is mediated by the binding of the BAG domain, which forms a three-helix bundle, to the N-terminal nucleotide-binding domain of Protein A. \synth{This coupling induces a conformational switch in the ATPase region of Protein A that is incompatible with nucleotide binding, thereby triggering the release of ADP and the subsequent release of substrate proteins from the central substrate-binding domain.} \synth{Protein B serves as a nucleotide exchange factor that specifically facilitates the conversion of Protein A from an ADP-bound state to an ATP-bound state, a critical step in regulating the chaperone's affinity for its polypeptides.}

$\dots$

Protein B acts as a co-chaperone that regulates the activity of Protein A while also exhibiting independent anti-apoptotic properties. \synth{It markedly increases the anti-cell death function of BCL2 in an ATP-dependent manner and inhibits the pro-apoptotic function of PPP1R15A.} $\dots$
\end{tcolorbox}
\caption{Example of allosteric coupling across named conformational states. Binding at one site (the N-terminal nucleotide-binding domain) propagates a conformational change to a physically distant site (the central substrate-binding domain), and the synthesis explicitly names the intermediate states (ADP-bound, ATP-bound, substrate-loaded, substrate-released).}
\label{fig:synth_hspa8_bag1}
\end{figure}

\begin{figure}[htbp]
\centering
\begin{tcolorbox}[ppiexamplebox,
  title={AXIN1 $\leftrightarrow$ GSK3$\beta$ (bistable switch in the Wnt destruction complex)}]
\small
Protein A and Protein B directly associate as core components of the beta-catenin destruction complex, a multimeric assembly that regulates the canonical Wnt signaling pathway by controlling the stability of cytoplasmic beta-catenin. Within this context, Protein B up-regulates Protein A via phosphorylation, a direct modification that targets the central region of Protein A at specific threonine and serine residues. \synth{This phosphorylation event is critical for maintaining Protein A in an active, open conformation, which enhances its ability to bind beta-catenin and facilitates the subsequent phosphorylation of beta-catenin by Protein B.} $\dots$

\synth{In the absence of Wnt signaling, the physical association between Protein A and Protein B promotes the hyperphosphorylation of beta-catenin, poising it for ubiquitination and proteasome-mediated degradation.} \synth{Upon Wnt stimulation, the destruction complex is recruited to the membrane, leading to the inhibition of Protein B and the subsequent nuclear accumulation of beta-catenin.} $\dots$
\end{tcolorbox}
\caption{Example of context-dependent bistable switch. The identical AXIN1$\leftrightarrow$GSK3$\beta$ physical association produces opposite cellular outcomes as a function of upstream pathway state: proteasomal destruction of $\beta$-catenin in the Wnt-off regime, and nuclear accumulation of $\beta$-catenin in the Wnt-on regime.}
\label{fig:synth_axin1_gsk3b}
\end{figure}

Our approach to modeling PPI with free-form description changes the paradigm. Instead of forcing an interaction into a predefined category, a free-form caption can describe the nature, context, mechanism and biological significance of an interaction with richer expressiveness and interpretability. Figures~\ref{fig:synth_synj1_amph}, \ref{fig:synth_tp53_mdm2}, \ref{fig:synth_hras_raf1}, \ref{fig:synth_hspa8_bag1}, and \ref{fig:synth_axin1_gsk3b} provide examples highlighting the power of natural language representations.

First, free-form text can capture nuances that categorical labels would ignore, including conditions for the interaction to occur, biological context, structural bias for binding, and relationships to disease and function. The ambiguity and uncertainty can also be effectively expressed in natural language. Standardized labeling in prior work leads to a significant bottleneck in extracting information from literature knowledge bases, whereas free-text synthesis enables more comprehensive and flexible summarization. 

Moreover, natural language descriptions are inherently human-readable, allowing for deeper exploration and generative reasoning. A free-text caption can propose mechanisms, suggest hypotheses, and explain why an interaction occurs, which are all beyond the capacity of classification or regression approaches. The flexible schema without predefined ontology categories facilitates the exploration of emerging interactions that were previously undiscovered. 

After all, the strength of PPI modeling lies not only in its ability to make predictions, ad also in how faithfully it can capture and express the complexity of the underlying biological reality.

\section{Data Construction}

\subsection{Sources of Raw Evidence}
\label{apd:10datasets}

The raw evidence of PPIs are gathered from ten comprehensive sources so that different aspects of the interaction are covered by different specialized datasets, as shown in Table \ref{tab:appendix-data-sources} with version and license information.

\begin{table}[htbp]
\centering
\caption{Sources for raw evidence. Feature columns are: Binary PPI (Bin), Complex/multi-subunit (Cpx), signed/causal Effect (Eff), Sequence (Seq), Functional/pathway/domain annotation (Fun), Literature (Lit), 3D Interface contact (Iface). PPI counts are pairs with non-empty data from the source, reported before and after quality-controlled filtering.}
\label{tab:appendix-data-sources}
\renewcommand{\arraystretch}{1.15}
\setlength{\tabcolsep}{4pt}
\resizebox{\textwidth}{!}{%
\begin{tabular}{l l l ccccccc r r}
\toprule
\textbf{Database} & \textbf{Version (retrieval)} & \textbf{License}
& \textbf{Bin} & \textbf{Cpx} & \textbf{Eff} & \textbf{Seq} & \textbf{Fun} & \textbf{Lit} & \textbf{Iface}
& \textbf{Before} & \textbf{After} \\
\midrule
IntAct        & rel.~2026-01-09\footnotemark[1]            & CC~BY~4.0      & $\bullet$ & $\bullet$ & --        & --        & --        & --        & $\bullet$ & $1{,}075{,}916$        & $351{,}515$ \\
PubMed        & live E-utilities (2026-04)\footnotemark[2] & NLM\,/\,public & --        & --        & --        & --        & --        & $\bullet$ & --        & $1{,}067{,}409$        & $350{,}224$ \\
UniProt       & rel.~2026\_01\footnotemark[3]              & CC~BY~4.0      & --        & --        & --        & $\bullet$ & $\bullet$ & --        & --        & $\phantom{0}997{,}480$ & $345{,}845$ \\
3did          & rel.~2024\_12\footnotemark[4]              & Academic       & --        & --        & --        & --        & --        & --        & $\bullet$ & $\phantom{00}77{,}200$ & $\phantom{0}68{,}736$ \\
Pfam          & rel.~37.0\footnotemark[5]                  & CC0~1.0        & --        & --        & --        & --        & $\bullet$ & --        & --        & $\phantom{0}770{,}243$ & $264{,}301$ \\
STRING        & v~12.0 (2025-07)\footnotemark[6]           & CC~BY~4.0      & $\bullet$ & --        & --        & --        & --        & --        & --        & $1{,}075{,}916$        & $351{,}515$ \\
SIGNOR        & v~4.0 (2026-04)\footnotemark[7]            & CC~BY-SA~4.0   & $\bullet$ & --        & $\bullet$ & --        & --        & --        & --        & $\phantom{000}4{,}737$ & $\phantom{00}4{,}737$ \\
Reactome      & v~96 (2026-04)\footnotemark[8]             & CC~BY~4.0      & --        & --        & --        & --        & $\bullet$ & --        & --        & $\phantom{0}783{,}894$ & $286{,}882$ \\
CORUM         & v~5.3 (2026-04)\footnotemark[9]            & CC~BY~4.0      & --        & $\bullet$ & --        & --        & --        & --        & --        & $\phantom{00}20{,}060$ & $\phantom{0}19{,}906$ \\
ComplexPortal & rel.~2026-01-09\footnotemark[10]           & CC0~1.0        & --        & $\bullet$ & --        & --        & --        & --        & --        & $\phantom{00}24{,}306$ & $\phantom{0}24{,}241$ \\
\bottomrule
\end{tabular}%
}
\end{table}
\footnotetext[1]{\url{https://ftp.ebi.ac.uk/pub/databases/intact/2026-01-09/}}
\footnotetext[2]{\url{https://eutils.ncbi.nlm.nih.gov/entrez/eutils/efetch.fcgi}}
\footnotetext[3]{\url{https://ftp.uniprot.org/pub/databases/uniprot/previous_major_releases/release-2025_04/}}
\footnotetext[4]{\url{https://3did.irbbarcelona.org/download/current/3did_flat.gz}}
\footnotetext[5]{\url{https://ftp.ebi.ac.uk/pub/databases/Pfam/releases/Pfam37.0/}}
\footnotetext[6]{\url{https://stringdb-downloads.org/download/v12.0/}}
\footnotetext[7]{\url{https://signor.uniroma2.it/releases/getLatestRelease.php}}
\footnotetext[8]{\url{https://reactome.org/download/current/UniProt2Reactome.txt}}
\footnotetext[9]{\url{https://mips.helmholtz-muenchen.de/fastapi-corum/public/file/download_current_file}}
\footnotetext[10]{\url{https://ftp.ebi.ac.uk/pub/databases/intact/complex/2026-01-09/}}

\subsection{Evidence Scoring}
\label{apd:evidence_score}

We assign each pair $r$ an \textbf{interaction} component evidence score $E_{\text{int}}(r)$ and a \textbf{context} component evidence score $E_{\text{ctx}}(r)$, and the overall scalar evidence score $E(r)$ is the gated aggregation of both:

\begin{align}
E_{\text{int}}(r) &= E_{\text{map}} + E_{\text{mech}} + E_{\text{lit}} + E_{\text{src}}, \quad\quad E_{\text{ctx}}(r) = \min\!\Bigl(\sum_{i \in \mathcal{C}} w_i\,\mathbbm{1}[c_i(r)],\; 4.0\Bigr), \\
E(r) &=
\begin{cases}
\min\bigl(E_{\text{int}}(r) + E_{\text{ctx}}(r),\; 5.0\bigr) & \text{if $r$ has no experimental detection method,} \\
E_{\text{int}}(r) + E_{\text{ctx}}(r) & \text{otherwise.}
\end{cases}
\end{align}

Per-signal weights and trigger conditions for the four interaction axes and the context axis are listed in Table~\ref{tab:appendix-evidence-score}.The scoring schema is designed/validated by human domain experts with proper biological motivations.

\renewcommand{\arraystretch}{1.0}
\setlength{\tabcolsep}{5pt}
\begin{longtable}{@{}l l l r@{}}
\caption{Evidence-score formula composed to comprehensively evaluate the richness of raw annotations in different aspects about the interaction.}
\label{tab:appendix-evidence-score}\\
\toprule
\textbf{Axis} & \textbf{Signal} & \textbf{Trigger} & \textbf{Weight} \\
\midrule
\endfirsthead
\multicolumn{4}{@{}l}{\textit{(Table~\ref{tab:appendix-evidence-score} continued)}}\\
\toprule
\textbf{Axis} & \textbf{Signal} & \textbf{Trigger} & \textbf{Weight} \\
\midrule
\endhead
\endfoot
\bottomrule
\endlastfoot

\multicolumn{4}{@{}l}{\textit{$E_{\text{map}}$ --- interaction type, interface mapping, named-complex grounding}}\\
& interaction type             & ``direct'' \textbf{else} ``physical association'' & $+2.0\,/\,+1.0$ \\
& binding-region features      & one side / both sides                            & $+1.0\,/\,+2.0$ \\
& 3did interface               & one side / both sides                            & $+1.0\,/\,+2.0$ \\
& named-complex reference      & non-empty                                        & $+2.0$ \\
& subunit mentions partner     & true                                             & $+1.0$ \\
\midrule

\multicolumn{4}{@{}l}{\textit{$E_{\text{mech}}$ --- mechanistic detail}}\\
& enzymatic interaction        & phospho / ubiq / cleav / acetyl / methyl / \ldots & $+1.0$ \\
& STRING action modes          & non-empty (\,$+\,$any score $\ge\!700$)          & $+1.0\,(+0.5)$ \\
& biophysical parameters       & non-empty                                        & $+1.0$ \\
& stoichiometry                & either side annotated                            & $+0.5$ \\
& biological roles             & either side, non-default                         & $+0.5$ \\
& self-interaction             & flag set                                         & $+0.5$ \\
\midrule

\multicolumn{4}{@{}l}{\textit{$E_{\text{lit}}$ --- literature and experimental redundancy}}\\
& publications                 & threshold crossings $\{2,\,5\}$                  & $+1.0$ each \\
& experimental method count    & $n_{\text{exp-fam}} \ge 2$                       & $+1.0$ \\
& IntAct miscore               & threshold crossings $\{0.45,\,0.56,\,0.65\}$     & $+0.5$ each \\
& evidence lines               & $\ge 5$                                          & $+0.5$ \\
& abstract length (chars)      & threshold crossings $\{1{,}500,\,3{,}000,\,5{,}000\}$ & $+1.0$ each \\
& interaction annotations      & $\ge 250$ chars                                  & $+0.5$ \\
& experimental detection       & non-computational (\,$+\,$not spoke-expanded)    & $+1.0\,(+0.5)$ \\
& shared pathways              & threshold crossings $\{1,\,3\}$                  & $+0.5$ each \\
& STRING combined score        & threshold crossings $\{400,\,700\}$              & $+0.25$ each \\
\midrule

\multicolumn{4}{@{}l}{\textit{$E_{\text{src}}$ --- curator-graded complex / mechanism sources}}\\
& SIGNOR                       & non-empty (\,$+\,$any entry tagged direct)       & $+2.0\,(+1.0)$ \\
& CORUM                        & non-empty                                        & $+2.0$ \\
& Complex Portal               & non-empty                                        & $+2.0$ \\
& CORUM $\wedge$ Complex Portal & cross-validated                                 & $+1.0$ \\
\midrule

\multicolumn{4}{@{}l}{\textit{$E_{\text{ctx}}$ --- contextual biological coherence (capped at $4.0$)}}\\
& paired UniProt fields        & both sides annotated: function, domains,                           & \\
&                              & \quad similarity...              & $+0.25$ each \\
& subcellular location         & shared term                                      & $+0.5$ \\
& shared GO term               & component / process / function                   & $+0.25$ each \\
& either-side annotation       & tissue, catalytic activity, PTM, modified                 & \\
&                              & \quad residues, regulation, active / binding             & \\
&                              & \quad  sites, motifs, Zn fingers, free-text, ...           & $+0.25$ each \\

& disease keywords             & either side (\,$+\,$both sides)                  & $+0.5\,(+0.25)$ \\
\end{longtable}

\subsection{K-Means Clustering}
\label{apd:kmeans}

\begin{figure} [thb!]
    \centering    \includegraphics[width=0.9\linewidth]{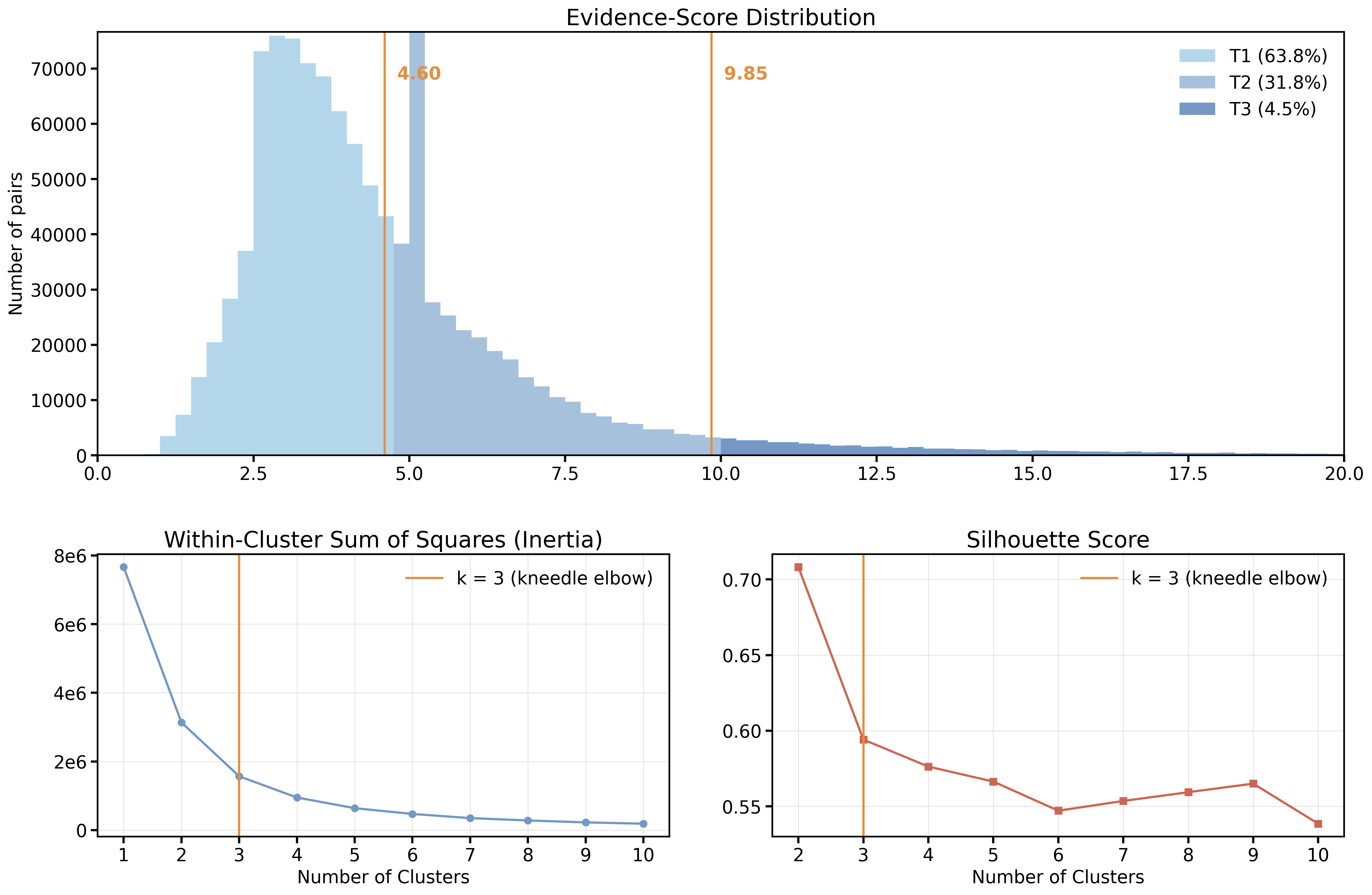}
    \caption{K-means clustering of the per-pair evidence score. Top: score distribution colored by three different clusters; bottom: inertia and silhouette scores for different numbers of clusters as well as the elbow k=3.}
    \label{fig:kmeans}
\end{figure}

We treat the per-pair evidence score as a 1D feature and run K-means clustering on all 1.08M pairs we gathered from sources of raw evidence. As shown in Figure~\ref{fig:kmeans}, different numbers of clusters are tested with both within-cluster sum of squares (inertia) and silhouette scores (probed with 200k subset) computed. The elbow with k=3 clusters is chosen from both curves with the kneedle criterion. The split criteria is consistent with our manual inspection, as most PPIs from T1 are mostly supported by one single high-throughput experiment and has no solid experimental evidence on its interaction details, as well as most PPIs from T3 are well supported by multiple single-sample experiments with strong evidence on interaction mechanism and biological effects. 

\subsection{Manifold Coverage and Filtering Bias}
\label{apd:bias}

\begin{figure} [thb!]
    \centering    \includegraphics[width=\linewidth]{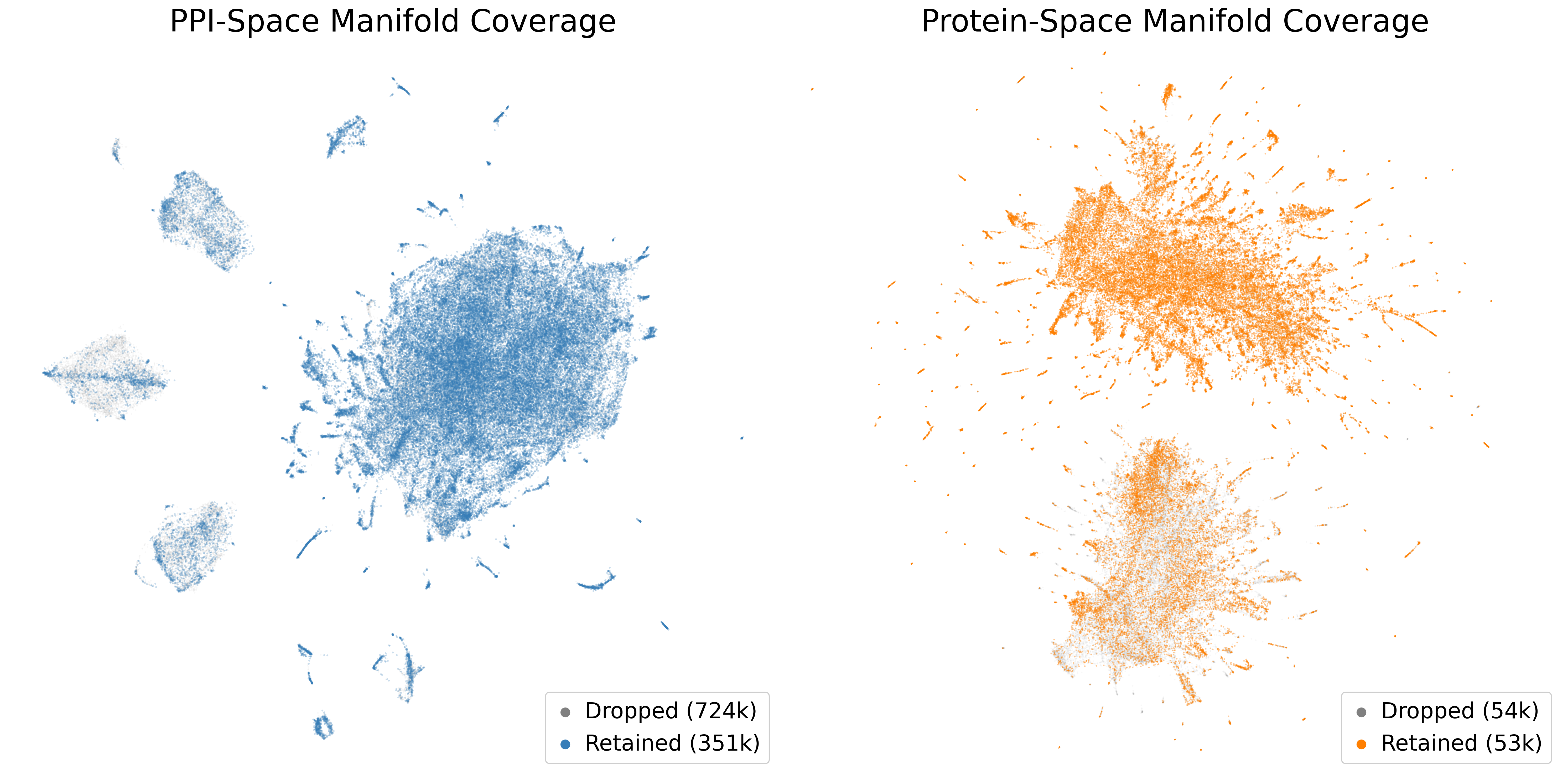}
    \caption{UMAP (Uniform manifold approximation and projection) visualization of manifold coverage of the filtered dataset against the original raw samples in both PPI space and protein space. The most of the original coverage is preserved.}
    \label{fig:umap_combined}
\end{figure}

With the help of K-means clustering, the dataset is filtered using an evidence-tiered quality control procedure. We analyze how this filtering affects diversity and coverage in the PPI description dataset. Specifically, we apply Principal Component Analysis (PCA) to project ESM3 protein embeddings to a lower-dimensional space. Figure~\ref{fig:umap_combined} compares the manifold coverage of the filtered dataset against the original raw samples in both PPI space and protein space. 

In the PPI space (left) where each data point represents a pair of proteins interacting, the filtered dataset closely overlaps with the high-density regions of the original distribution, indicating that the dominant interaction modes are well preserved. While a small number of peripheral clusters and low-density regions are reduced, the overall geometric structure of the manifold remains intact. 

And in the protein space (right), we observe a similar trend. The filtered dataset retains nearly all high-density regions, with only sparse outliers and boundary regions being pruned. Notably, the filtering appears to preferentially remove isolated or weakly supported samples rather than systematically biasing the representation toward specific regions of the space. 

\begin{figure} [thb!]
    \centering
    \includegraphics[width=0.7\linewidth]{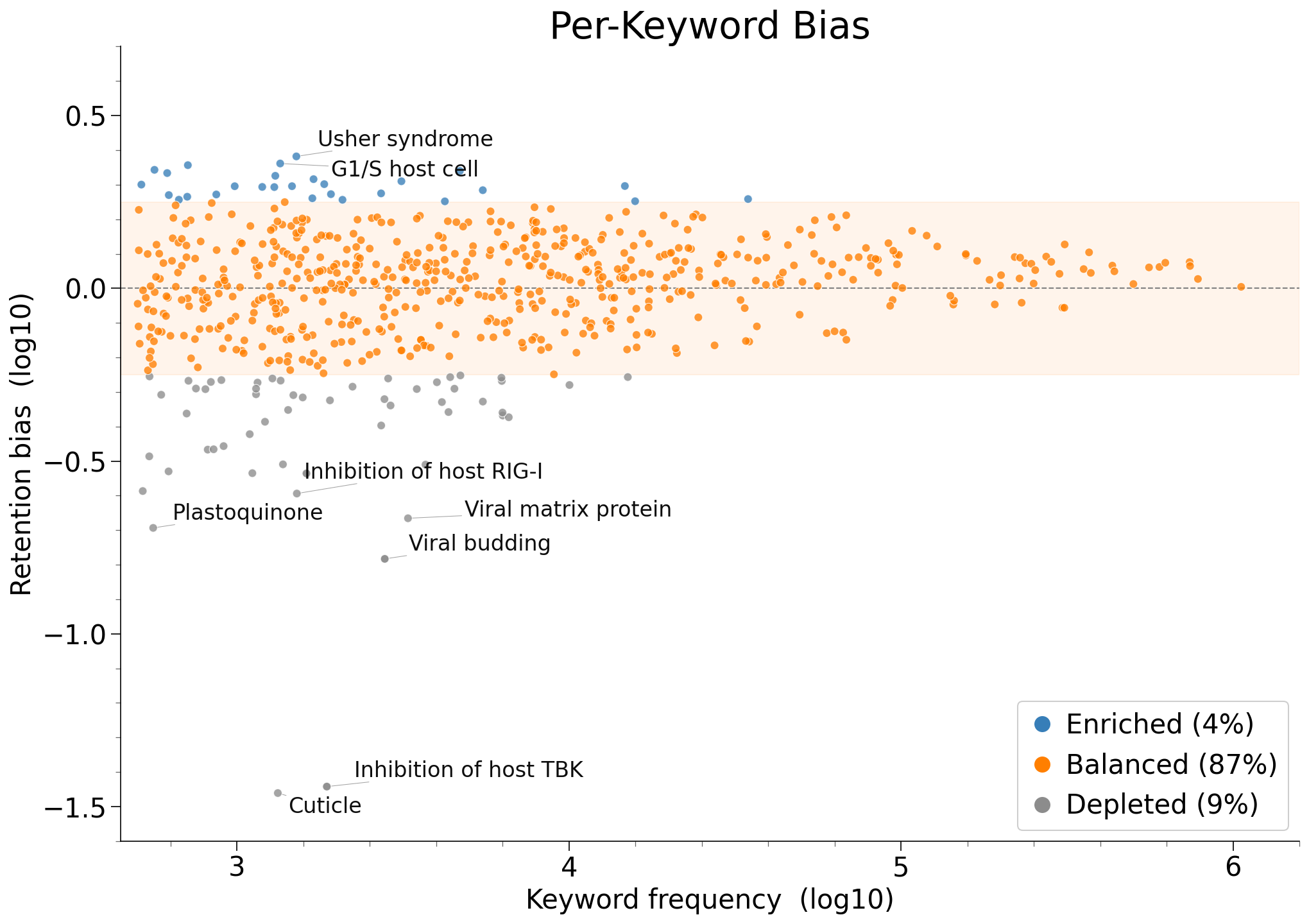}
    \caption{Retention bias as a function of keyword frequency. Most high-frequent keywords in the raw dataset are preserved in the filtering.}
    \label{fig:bias_scatter_keyword}
\end{figure}

We further examine whether the filtering procedure introduces bias at the semantic level by analyzing retention rates across annotation keywords. Figure~\ref{fig:bias_scatter_keyword} plots retention bias as a function of keyword frequency. The majority of keywords remain balanced, with retention rated concentrated near zero across the full frequency spectrum, indicating that the filtering process does not systematically favor or suppress most functional annotations. 

A small fraction of keywords (4\%) are enriched, primarily corresponding to well-supported or frequently studies biological contexts, while 9\% are depleted, often associated with more specific or sparsely represented mechanisms especially for virus. Notably, the depletion mostly happens to lower-frequency keywords, suggesting that the filtering preferentially removes under-supported or noisier annotations.  

Overall, these results indicate that the evidence-tiered filtering preserves the global semantic distribution of the dataset while selectively pruning less reliable or weakly supported annotations, without introducing substantial bias. 

\subsection{Evidence-tiered Prompting}
\label{apd:tags}

For the augment of the PPIs with an LLM process, we introduce a tier-controlled generation framework that explicitly constrains the output space according to the strength of biological support. Four orthogonal constraints are explicitly enforced in each prompt: \textit{descriptive granularity}, \textit{epistemic strength}, \textit{mechanistic attribution}, and \textit{silence policy} further described as follows:

\begin{itemize}
    \item First, we regulate \textbf{descriptive granularity} through tier-specific length constraints, so that low-confidence interactions are restricted to concise descriptions while high-confidence PPIs are allowed with more detailed interpretation. 

    \item Second, we control the \textbf{epistemic strength} of language via tier-conditioned verb selections. Interactions with strong mechanistic support are instructed to use assertive biological terms, whereas weakly supported interactions are required to use hedged expressions to reflect observational evidence. 

    \item Third, we regulate the \textbf{mechanistic attribution} by conditioning generation on the availability of structural or curated mechanistic annotations. When such evidence is absent, the teacher model is explicitly constrained from introducing molecular-level mechanistic explanations to prevent unsupported speculation. 

    \item Fourth, we enforce a \textbf{silence policy} on under-annotated entities to restrain over-description of proteins beyond available evidence, avoiding implicit hallucination through prior-driven completion. 
\end{itemize}

\begin{figure} [thb!]
    \centering
    \includegraphics[width=\linewidth]{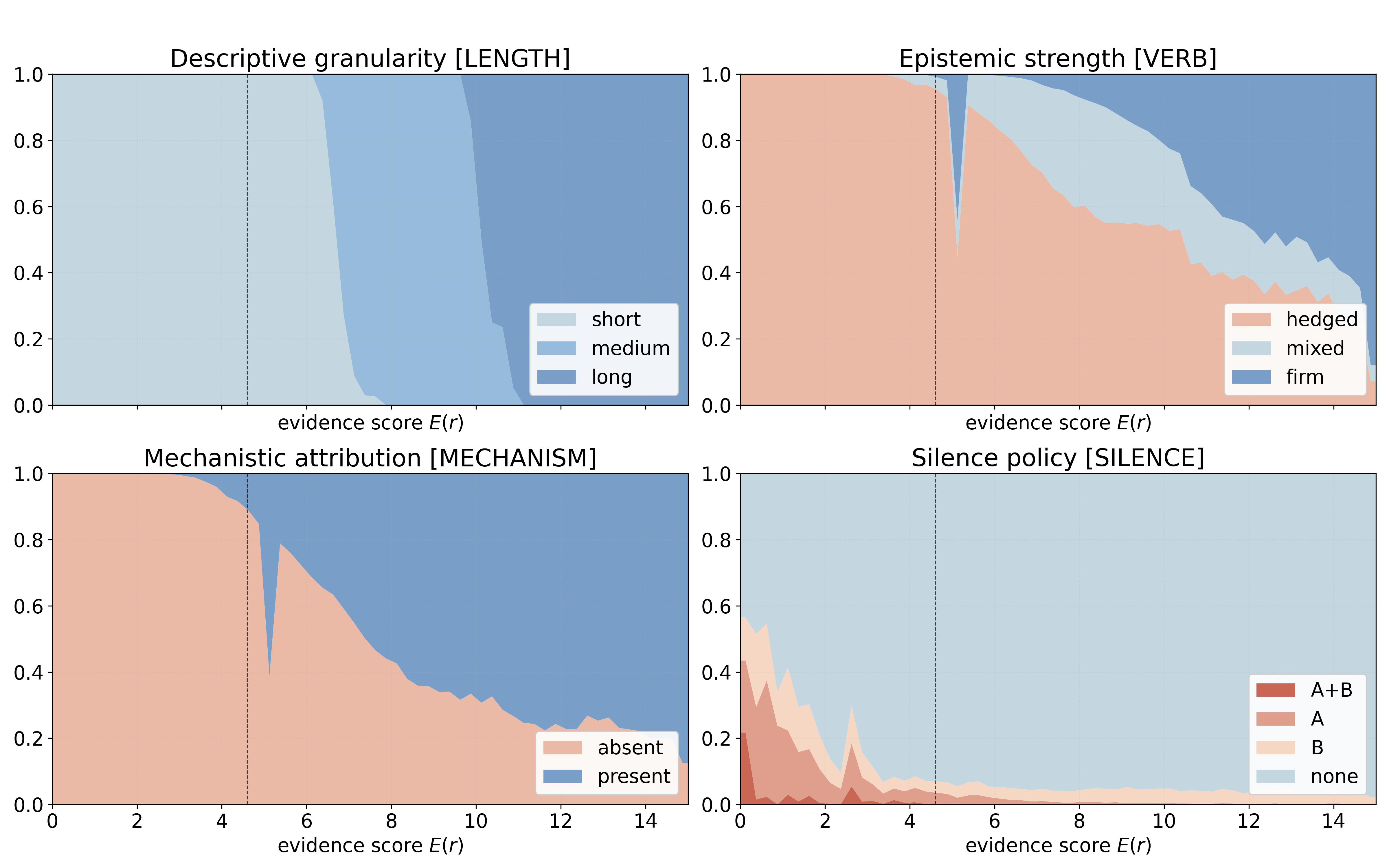}
    \caption{Statistics of the augmented PPI dataset samples with the proposed evidence-tiered constraints. A substantial variability can be observed across PPIs, making tier-controlled dynamic prompting necessary for synthesis.}
    \label{fig:tags}
\end{figure}

We show some statistics related to the dataset components after the augmentations to each of the constraints in Figure \ref{fig:tags}. 

\section{ESM3 as a Single-Protein Encoder}
\label{ap:ESM3}

The main paper assumes a sequence-only setting, where PPI2Text requires only amino acid sequences as the only input. However, ESM3 is not inherently a sequence-only encoder, but rather a generative multimodal model that expects multiple per-residue input tracks, including structural information. This mismatch by design is addressed by reconstructing all required ESM3 inputs from the protein sequence alone using external predictors, and then perform a frozen forward pass to obtain per-residue embeddings.

For a protein of length $L$, we assemble the full set of per-residue inputs expected by ESM3. All inputs are ultimately derived from the amino acid sequence, ensuring consistency with the sequence-only assumption.

\begin{itemize}
  \item \textbf{Sequence tokens:}  
  The amino acid sequence is retrieved from UniProt and tokenized using the ESM3 vocabulary, including BOS and EOS tokens, resulting in a sequence of length $L+2$.

  \item \textbf{Backbone coordinates:}  
  We obtain the sequence-based predicted 3D structure from AlphaFold2. From this structure, we extract the backbone atom coordinates (N, C$_\alpha$, C) for each residue, which are required by ESM3’s geometric attention module.

  \item \textbf{Discrete structure tokens:}  
  The backbone coordinates are passed through the frozen ESM3 structure encoder (a Vector Quantized Variational Autoencoder (VQ-VAE) released with ESM3) to produce one discrete structure token per residue.

  \item \textbf{Secondary structure (SS8):}  
  We compute 8-class secondary structure assignments using the mkdssp library applied to the AlphaFold structure. To align with the ESM3 token set, DSSP symbols \texttt{-} (coil) and \texttt{P} (polyproline-II) are both mapped to \texttt{C} (cysteine).

  \item \textbf{Solvent-accessible surface area (SASA):}  
  Residue-level SASA values are computed using the Shrake--Rupley implementation in the BioPython library, and then
  discretized using the ESM3 SASA tokenizer.

  \item \textbf{Confidence scores (pLDDT):}  
  Per-residue confidence scores are read from the AlphaFold PDB B-factor column and normalized to $[0,1]$ by dividing by 100. We also compute the mean pLDDT across residues as a global confidence estimate.

  \item \textbf{Function and residue annotation tracks:}  
  ESM3 expects additional tracks corresponding to functional annotations (e.g., InterPro) and free-form residue annotations. To avoid potential information leakage affecting the evaluation process, both fields are not provided but filled with their respective padding tokens instead.
\end{itemize}

Although several inputs depend on structural information, all such features are derived from auxiliary predictions, which are generated from sequence. Thus, the overall pipeline adheres to the sequence-only constraint.

The assembled inputs are fed into ESM3 using a single chain identifier. The core computation of ESM3 are 48 transformer layers stacked with geometric attention. We retain only the final hidden representation corresponding to the post-LayerNorm activations of the last transformer block. All other outputs, including decoder heads and structure/function predictions, are all discarded.

\section{PaCo-RoPE}
\label{ap:PaCo-R}

\subsection{Formulation}

We formalize PaCo-RoPE as an extension of rotary positional encoding (RoPE) to 3D position indices with channel interleaving to encode single protein representations and the interaction pair map simultane. 

For a query/key vector $x \in \mathbb{R}^{d_h}$, a standard 1D RoPE applies a rotation to each frequency pair $(2i, 2i+1)$:

\begin{equation}
\mathrm{RoPE}(x, p)_{
\{2i,2i+1\}
}
=
\begin{pmatrix}
\cos \theta_i(p) & -\sin \theta_i(p) \\
\sin \theta_i(p) & \cos \theta_i(p)
\end{pmatrix}
x_{\{2i,2i+1\}},
\quad
\theta_i(p) = \omega_i \, p,
\end{equation}

where $\omega_i = \omega_0^{\,2i/d_h}$ is the frequency.

Each token is assigned a triplet $\mathbf{p} = (p^T, p^\theta, p^\varphi) \in \mathbb{R}^3$ depending on its modality and semantic role:

\begin{itemize}
    \item \textbf{Text tokens.}  
    For a token at sequence position $n$,
    \begin{equation}
        \mathbf{p} = (n, n, n).
    \end{equation}

    \item \textbf{Single-protein tokens (protein A).}  
    For a token at sequence position $n$, representing residues centered at index $i$ in protein $A$,
    \begin{equation}
    \mathbf{p} = \bigl(n, \lfloor i / s \rfloor, 0\bigr).
    \end{equation}

    \item \textbf{Single-protein tokens (protein B).}  
    For a token at sequence position $n$, representing residues centered at index $j$ in protein $B$,
    \begin{equation}
    \mathbf{p} = \bigl(n, 0, \lfloor j / s \rfloor\bigr).
    \end{equation}

    \item \textbf{Pair-map tokens.}  
    Let a pair-map be partitioned into a grid of size $H \times W$. For a patch at grid location $(m, n)$ (zero-indexed), serialized at sequence position $k$, we define
    \begin{equation}
    \mathbf{p}
    =
    \left(
    k,\,
    \left(m+\tfrac{1}{2}\right)\frac{L_A'}{H},\,
    \left(n+\tfrac{1}{2}\right)\frac{L_B'}{W}
    \right),
    \end{equation}
    where $L_A'=\lfloor\frac{L_A}{4}\rfloor$ and $L_B'=\lfloor\frac{L_B}{4}\rfloor$ denote the effective lengths of proteins $A$ and $B$ after 1D compression, and the $+\tfrac{1}{2}$ term centers each patch within its spatial bin.
\end{itemize}

Here $s$ is the stride used to compress residue indices to token indices. All coordinates are real-valued, enabling seamless integration of discretized sequence tokens and pooled pair-map patches.

Let $i \in \{0, \dots, d_h/2 - 1\}$ index frequency pairs. We define a channel selector
\begin{equation}
c(i) =
\begin{cases}
T, & i \bmod 3 = 0, \\
\theta, & i \bmod 3 = 1, \\
\varphi, & i \bmod 3 = 2.
\end{cases}
\end{equation}
Each frequency pair is thus assigned to exactly one positional channel. Unlike the block partitioning M-RoPE, the interleaving scheme keeps any single channel from dominating a contiguous frequency band.

So that for each pair $(2i, 2i+1)$, we apply PaCo-RoPE using the selected coordinate:
\begin{equation}
\theta_i(\mathbf{p}) = \omega_i \cdot p^{\,c(i)},
\end{equation}
\begin{equation}
\mathrm{PaCoRoPE}(x, \mathbf{p})_{\{2i,2i+1\}}
=
\begin{pmatrix}
\cos \theta_i(\mathbf{p}) & -\sin \theta_i(\mathbf{p}) \\
\sin \theta_i(\mathbf{p}) & \cos \theta_i(\mathbf{p})
\end{pmatrix}
x_{\{2i,2i+1\}}.
\end{equation}

\subsection{Interpretation}

Consider protein A as an example. Let protein A consist of $L_A'$ effective residue-level tokens after tokenization, indexed by
\begin{equation}
r \in \{0, \dots, L_A' - 1\}.
\end{equation}
Protein-A tokens are mapped to positional coordinates via
\begin{equation}
p^\theta(r) = \left\lfloor \frac{r}{s} \right\rfloor,
\end{equation}
where $s$ is the token stride.

The pair-map is constructed by partitioning the same residue axis into $H$ bins using average pooling. Define the binning function
\begin{equation}
\pi_H(r) = \left\lfloor \frac{r}{L_A'/H} \right\rfloor.
\end{equation}
Each pair-map row index $m \in \{0,\dots,H-1\}$ corresponds to the residue interval
\begin{equation}
r \in \left[m \cdot \frac{L_A'}{H}, (m+1)\cdot \frac{L_A'}{H}\right),
\end{equation}
and is assigned the centered coordinate
\begin{equation}
p^\theta_{\text{pair}}(m)
=
\left(m + \tfrac{1}{2}\right)\frac{L_A'}{H}.
\end{equation}

For any residue $r$ such that $m = \pi_H(r)$, we have
\begin{equation}
\left| r - p^\theta_{\text{pair}}(m) \right| \le \frac{L_A'}{2H}.
\end{equation}
Thus, protein-A tokens and pair-map tokens share a common discretization of the same underlying residue axis, up to bounded quantization error $\mathcal{O}(L_A'/H)$.

An identical construction applies to protein B, yielding the same alignment property along the second spatial axis: 
\begin{equation}
\left| r - p^\varphi_{\text{pair}}(n) \right| \le \frac{L_B'}{2W}.
\end{equation}

Therefore, protein-$B$ tokens and pair-map tokens share the same partitioning of the residue axis up to bounded quantization error $\mathcal{O}(L_B'/W)$.

Thus, PaCo-RoPE does not impose exact equality between token and pair-map coordinates. Instead, both representations are derived from a shared discretization of the same underlying residue axis, ensuring consistent spatial correspondence under bounded resolution error induced by pooling and tokenization.

\section{Hyperparameters and Computational Resources} 
\label{apd:hparams}

We use \texttt{esm3-sm-open-v1} as the single protein encoder with 1.4B parameters and 1536 as the hidden dimension. Due to the limitation of the encoder, PPIs with proteins no longer than 2048 amino-acids are kept. Both stride-2 Conv1D layers use kernel of size 4, resulting in a four times length reduction with the pipeline hidden dimension of 1024. The compression factor of 4 corresponds to the average size of $alpha$-helical turns, and the receptive field of 7 covers the natural length scale of local secondary-structure elements. The adaptive mean pooling targets a $32\times 32$ grid, resulting in 1024 pair map tokens to the decoder, as a trade-off between information granularity and computation efficiency. We choose \texttt{Qwen3-4B-Instruct} as the frozen base decoder with excellent generic knowledge in biology. LoRA adapter of rank 32 and 0.1 dropout rate is applied to provide enough flexibility and room for the new knowledge to be injected through supervised fine-tuning (SFT). The end-to-end SFT takes 200 GPU hours on NVIDIA H100 80GB for 5 epochs with effective batch size of 64. The peak learning rate for LoRA components is 1e-4, and 5e-5 for modules initialized from scratch, both managed by warm-up and cosine scheduling.

\section{Dataset Splitting Protocol and Experiments}
\label{ap:SplittingProtocol}

\paragraph{Temporal holdout:} To evaluate performance of our model in a realistic prospective setting, we introduce a temporal split at May 1, 2025, so that all PPIs first annotated after this date form the holdout set, simulating prediction on newly discovered interactions. To prevent information leakage from homologous proteins, we perform sequence-based decontamination using MMSeq2, defining two proteins A and B as similar ($A\sim B$) if they share more than 50\% sequence identity over at least 80\% coverage. For the holdout setting, the decontamination is imposed at the pair level: a training PPI $(A,B)$ is removed if there exists a test PPI $(C,D)$ that: 

\begin{equation}
    \Big((A\sim C) \cap (B\sim D)\Big) \cup \Big((A\sim D) \cap (B\sim C)\Big)
\end{equation}

\paragraph{C3-hard:} 

Following the idea of Bernett (\cite{10.1093/bib/bbae076}), we construct a more stringent C3-hard test set to evaluate out-of-distribution generalization. The leakage criterion is defined at the single-protein level, such that a training interaction $(A,B)$ is excluded if there exists a test interaction $(C,D)$ satisfying:

\begin{equation}
    (A\sim C)\cup(A\sim D)\cup(B\sim C)\cup(B\sim D)
\end{equation}

This ensures that no homologous proteins are shared between the training and test sets, effectively removing any opportunity for the model to exploit sequence similarity or evolutionary relatedness. As a result, C3-hard represents a highly controlled and intentionally stringent evaluation setting that stresses compositional generalization beyond realistic application scenarios.

While this setting is valuable as a worst-case robustness test, it is more extreme than typical pharmaceutical and wet-lab discovery conditions, where newly studied proteins often retain varying degrees of sequence or functional similarity to previously characterized ones and such biological priors are actively leveraged in practice. Consequently, C3-hard should be interpreted as an upper-bound stress test rather than a direct proxy for standard real-world use cases. In practice, this setting is expected to significantly increase task difficulty for all methods, while more closely reflecting worst-case extrapolation behavior compared to more realistic in-distribution or temporally shifted evaluation settings.

In addition to Figure \ref{fig:graphs}, Table \ref{tab:quant-ppi2text-C3} provides a more detailed comparison of the baseline model and ablation variants on the proposed C3-hard split. The results follow the same overall trend observed in Table \ref{tab:quant-ppi2text-hold}. Specifically, removing the encoder leads to the weakest performance across all settings, with a particular degradation in the raw evidence-based metrics (see Appendix \ref{ap:raw_eval_prompt} for further details on these metrics). In contrast, the best performance is consistently achieved by configurations that incorporate all three embedding components, with the full PPI2Text model obtaining the highest scores across the majority of evaluation metrics.

  \begin{table}[htbp]
  \small
  \centering
  \caption{Baseline and ablation results on C3-hard split. Lexical metrics includes BLEU-2/4 F1 scores (B-2/4), ROUGE-1/2/L F1 scores (R-1/2/L), with semantic BERTScore using RoBERTa (RBT) and BioBERT (BBT) embeddings. LLM-as-a-judge scores against the raw evidence cover Entities (Ent), Interaction (Int), Mechanism (Mec), and the Average (Avg).}
  \resizebox{\textwidth}{!}{%
  \begin{tabular}{l!{\vrule width 1pt}cc!{\vrule width 0.5pt}ccc!{\vrule width 0.5pt}cc!{\vrule width 1pt}ccc!{\vrule
   width 0.5pt}c}

  \toprule

  \multirow[b]{2}{*}{\textbf{Model}} & \multicolumn{7}{c!{\vrule width 1pt}}{\textbf{(against Synthesized Text)}} &
  \multicolumn{4}{c}{\textbf{(against Raw Evidence)}} \\
  [4pt]

   & \textbf{B-2} & \textbf{B-4} & \textbf{R-1} & \textbf{R-2} & \textbf{R-L} & \textbf{RBT} & \textbf{BBT} &
  \textbf{Ent} & \textbf{Int} & \textbf{Mec} & \textbf{Avg} \\

  \midrule

  Seq+Qwen3            & 34.16          & 17.82          & 51.16          & 20.11          & 26.60
    & 87.11          & 80.26          & 1.45          & 4.71          & 0.38          & 2.18          \\
  MINT+Qwen3                    & 34.87          & 18.47          & 51.71          & 22.23          & 27.53
    & 87.32          & 81.10          & 2.53          & 5.68          & 1.50          & 3.24          \\
  \midrule
  SingProt-Only PPI2Text                    & 34.87          & 18.74          & 52.81          & 23.11          & 28.58
    & 87.72          & 81.52          & 3.01          & 5.99          & 1.24          & 3.41          \\
  PairMap-Only PPI2Text           & 35.35          & 19.26          & 53.26          & 23.98          & 28.81
    & 87.93          & 81.73          & 4.13          & 6.49          & 2.25          & 4.29          \\
  1D-RoPE PPI2Text                & 34.99          & 19.00          & 52.81          & 23.61          & 28.69
    & 87.72          & 81.10          & 2.42          & 6.18          & 2.63          & 3.74          \\
  No-Cross PPI2Text               & 35.35          & 18.87          & 52.48          & 23.23          & 28.23
    & 87.52          & 80.89          & 4.02          & 7.27          & 4.58          & 5.29          \\
  \midrule
  \textbf{PPI2Text}               & \textbf{37.26} & \textbf{20.57} & \textbf{55.02} & \textbf{25.48} &
  \textbf{29.74} & \textbf{88.23} & \textbf{82.57} & \textbf{5.73} & \textbf{7.88} & \textbf{5.33} & \textbf{6.31} \\
  \bottomrule
  \end{tabular}
  }
  \label{tab:quant-ppi2text-C3}
  \end{table}

\section{Evaluation Against Raw Evidence: LLM-as-a-Judge}
\label{ap:raw_eval_prompt}

Linguistic metrics evaluated on generated text against the synthesized reference text are well established and widely accepted evaluation protocol, but the augmented nature of the reference text adds a layer of uncertainty. Despite careful examination by human expert biologists on the quality of the synthesized reference text, we cannot rule out edge cases where minor details are misinterpreted or missing. In this case, we propose to evaluate the prediction against the aggregated raw evidence for each PPI, since the annotation from human-curated sources are well accepted as gold-standard ground-truth. 

To enable large-scale evaluation of generated predictions, relying solely on biochemical experts would be time consuming, as validating thousands of outputs requires substantial manual effort. To address this limitation, we propose the use of a large language model (Claude-Opus-4.7) as an automated judge, enabling a faster and more scalable evaluation of our pipeline. The prompt instructs the LLM to act as an expert biochemist and systematically compare the predicted interaction description against the evidence card along four orthogonal evaluation axes. 

First, entity grounding measures whether the prediction correctly identifies proteins, families, structural features, organism, and cellular context without introducing unsupported or fabricated details. This dimension is critical because errors at the level of entity identity or biological context fundamentally undermine the validity of any downstream interpretation. 

Second, interaction topology evaluates whether the predicted relationship type (e.g., direct binding, complex co-membership, or co-occurrence) matches the level of evidence supported by curated databases such as UniProt and STRING (Appendix \ref{apd:10datasets}), penalizing both over- and under-interpretation. Accurately capturing interaction topology ensures that the model's predictions respect the evidence strength and biological meaning encoded in source databases.

Third, mechanism fidelity assesses whether mechanistic details, such as directionality, regulatory effects, or domain-level interfaces, are accurately preserved when present in the underlying evidence. Finally, the framework accounts for cases where mechanistic information is absent, ensuring that models are not rewarded for fabricating unsupported mechanistic claims. Incorrect mechanistic predictions underlies its utility for explaining biological processes and guiding experimental design. Distortion of such details reduces rich biochemical knowledge to vague associations, while incorrect mechanisms can lead to misleading hypotheses about regulation or function.

Each axis is scored independently on a 0–10 scale, enabling a fine-grained decomposition of model performance across factual correctness, interaction classification, and mechanistic reasoning. In Figure \ref{fig:eval-prompt_raw} we provide the curated evaluation prompt. 

\begin{tcolorbox}[
  colback=gray!5,
  colframe=black,
  boxrule=0.8pt,
  arc=2mm,
  width=\linewidth
]
\small

You are an expert biochemist and protein-protein interaction reviewer. You are given (1) a compact RAW EVIDENCE CARD describing what is known about a protein-protein interaction from primary databases (UniProt, IntAct, STRING, Reactome, 3DID, SIGNOR, CORUM, ComplexPortal, PubMed abstracts) and (2) a model PREDICTION — a free-text description of that same interaction.\\[2pt]
Your job is to score the PREDICTION against the RAW EVIDENCE on THREE ORTHOGONAL AXES, each on an integer scale 0-10:\\[2pt]
\textbf{1. ENTITY GROUNDING —} Are the proteins, families, Pfam/structural folds, organism, subcellular localization, and individual protein functions described in the prediction consistent with the evidence card? Penalize fabricated families/folds, wrong organism, confidently wrong substrate names, invented cell-line context not in the abstracts.\\[2pt]
\textbf{2. INTERACTION TOPOLOGY —} Does the prediction correctly classify the *type* of association? Categories: self/homo-oligomer; direct binary binding; complex co-membership (named complex); database-level co-occurrence (single-report proteomic survey, no defined coupling). Award high scores for matching the truth's category and naming the right complex (when one exists). Penalize topology inflation (calling a DB-level co-occurrence a direct binding) or topology deflation (calling a known complex member just co-localized).\\[2pt]
\textbf{3. MECHANISM FIDELITY —} If the evidence carries mechanistic content (interface\_summary, causal\_effect, named complex membership requirement, mech\_detail $=$ present, 3DID Pfam pair), does the prediction preserve it: directional action (A acts on B), regulatory consequence, interface region/domain pair? Award 0-3 for flattening rich mechanism into "uncharacterized"; 4-6 for partial preservation; 7-10 for faithful capture. Use "NA" if the truth has no mechanism (mech\_detail $=$ absent and no interface\_summary and no causal\_effect) AND the prediction also avoids inventing one. If the truth has no mechanism but the prediction fabricates one, score LOW (don't use NA).\\[2pt]
Output in a JSON with three scores plus a one-sentence note (about 25 words).
\end{tcolorbox}
\noindent\begin{minipage}{\textwidth}
\captionof{figure}{Structured prompt that instructs the LLM to act as an expert biochemist and compare free-text interaction predictions against aggregated raw evidence.}\label{fig:eval-prompt_raw}
\end{minipage}

\section{Safeguards}
\label{sec:safe}

PPI2Text demonstrates strong potential for protein–protein interaction description, though we cannot rule out the possibility of occasional inaccuracies, so results should be interpreted with care and verified where possible. While the model is designed for beneficial research applications, we encourage its use in responsible and well-regulated settings to ensure alignment with ethical and safety standards.

\newpage

\section{Gemini Synthesis System Prompt}
\label{ap:synthesisPrompt}

\begin{tcolorbox}[
  colback=gray!5,
  colframe=black,
  boxrule=0.8pt,
  arc=2mm,
  width=\linewidth
]
\small

You are rewriting noisy, multi-source metadata about a protein-protein interaction into a single coherent scientific paragraph. The user prompt is organised in two halves:\\[2pt]
(1) DECISION BLOCKS — short lines in the form `[TAG] value` at the top of the user prompt. They encode pre-computed decisions that have already taken account of the evidence. Treat them as binding constraints. Do not re-derive, override, or question them.\\[2pt]
(2) EVIDENCE BLOCKS — labelled multi-source raw evidence about the interaction. These are the factual substrate of the paragraph.\\[4pt]
\textbf{DECISION BLOCK REFERENCE} \\[2pt]
\textbf{[Length]} e.g. "240-350 words, approximately 9-13 sentences" — the paragraph MUST land inside this range. Stop when the evidence has been covered; do not pad to hit the maximum, and do not truncate to stay under the minimum. If you finish with more to say than the target allows, prioritize interface / mechanism / complex context over single-protein housekeeping.\\[2pt]
\textbf{[Verb]} firm | mixed | hedged — governs the verbs you may use.\\
- firm: "directly binds", "the interface lies between", "directly contacts". No hedges in the main claim.\\
- mixed: "binds", "contacts", "physically associates". Mild hedging allowed for specific sub-claims only.\\
- hedged : "reported to co-purify with", "associates with", "appears to co-occur". Never use firm verbs.\\[2pt]
\textbf{[Mechanism]} present | absent — when absent, describe each protein individually and label the link as indirect, computational, or database-level. Do NOT invent a specific coupling mechanism in this case (no "B recruits A to …", no "their interaction streamlines X").\\[2pt]
\textbf{[Silence]} none | A | B | A+B — if a side is listed, do not assign that protein any functional descriptor. You may state its structural features (domains, localisation) plainly, but not "functions as", "serves as", "acts as", "plays a role in". Its family name, UniProt keywords, and protein-class descriptor are also forbidden as functional glosses for a silenced protein.\\[4pt]
\textbf{OUTPUT}\\[2pt]
One single paragraph of third-person scientific prose. No bullets, headers, markdown, preamble, trailing commentary, or quotation. The opening sentence must reflect whatever is most informative about THIS specific pair — the interface, the named complex, the asymmetry in function between the two proteins, the mechanism of the coupling, the strength or absence of evidence, or the location pattern. \\[4pt]
\textbf{BINDING RULES}\\[2pt]
1. Factual accuracy. Every protein-specific claim (partners, diseases, PTMs, activities, domains, tissues, clinical relevance) must come directly from an evidence block. You may add generic biochemical reasoning, but do not invent protein-specific details. If the evidence is silent, omit it.\\[2pt]
2. Naming and numbers. Do not name any specific experimental assay, technique, or acronym. This applies to anything listed under "Detection methods" or mentioned by name in an abstract (e.g., affinity-pulldowns, hybrid systems, mass-spectrometric surveys, array platforms, structural methods, complementation assays, proximity-labelling, surface or kinetic assays, kit or platform names). Use only generic terms ("biochemical evidence", "proteomic survey", "binding assay", "structural analysis", "high-throughput screen", "interactome data") or omit methods. \\[2pt]
3. Speculative-mechanism ban. Do not introduce regulation, competition, displacement, rescue, recycling, sequestration, or compartment-switching for this pair unless explicitly supported by an evidence block. You may describe each protein’s regulation, but not combine them into a specific mechanism without evidence.\\[2pt]
4. PTM specificity. Use PTM terms exactly as given in the evidence ("ubiquitinated", "neddylated", "SUMOylated"). Do not convert the generic "Ubl conjugation" into a specific modifier unless the evidence specifies it.\\[2pt]
5. Topology vocabulary. Use the provided qualitative position terms ("N-terminal", "central", "C-terminal", etc.) exactly as written. Do not use residue numbers or invent positions.\\[2pt]
6. Naming of the two proteins. Refer to them only as "Protein A" and "Protein B". Do not include real names, gene symbols, accessions, or species synonyms.\\[2pt]
7. Stop when the evidence has been covered. Do not add closing remarks about unknowns, future work, or unproven consequences. End with the last supported claim.\\[4pt]
Return ONLY the paragraph.
\end{tcolorbox}

\noindent\begin{minipage}{\textwidth}
\captionof{figure}{System prompt used to synthesize the free-text descriptions. A decision block reference helps the model to interpret evidence-tiered dynamic controlling labels.}\label{fig:gemini_prompt}
\end{minipage}



\end{document}